\documentclass[graybox]{svmult}

\usepackage{mathptmx}       
\usepackage{helvet}         
\usepackage{courier}        
\usepackage{type1cm}        
\usepackage{makeidx}         
\usepackage{graphicx}        
\usepackage{multicol}        
\usepackage[bottom]{footmisc}
\usepackage[numbers,sort]{natbib}

\makeindex             

\def\aap{A \& A}

\def\aj{AJ}
\def\apj{ApJ}
\def\apjl{ApJL}

\def\apjs{ApJS}

\def\araa{ARAA}
\def\mnras{MNRAS}
\def\nat{Nature}

\def\prd{PhRD}

\newcommand{\blya}{b_{\rm IGM}}
\newcommand{\tlya}{T_{\rm IGM}}
\def \dhobject{SDSS1558-0031}
\newcommand{\oicm}{\Omega_{\rm ICM}}
\newcommand{\olya}{\Omega_{\rm Ly\alpha}}
\newcommand{\oovi}{\Omega_{\rm OVI}}
\newcommand{\omg}{$\Omega_{neut}$}
\newcommand{\momg}{\Omega_{neut}}

\newcommand{\mohtwo}{\Omega_{\rm H_2}}

\newcommand{\kms}{km~s$^{-1}$}
\newcommand{\lya}{Ly$\alpha$}
\newcommand{\lyb}{Ly$\beta$}
\def\intl{\int\limits}

\newcommand{\mmsun}{M_{\odot}}

\def\h2{H$_2$}
\def\f0{$F_0$}

\newcommand{\cm}[1]{\, {\rm cm^{#1}}}
\newcommand{\mkms}{{\rm \; km\;s^{-1}}}

\newcommand{\N}[1]{{N({\rm #1})}}
\newcommand{\sci}[1]{{\rm \; \times \; 10^{#1}}}
\newcommand{\mnhi}{N_{\rm HI}}

\def\fnhi{$f(\mnhi,X)$}
\def\mfnhi{f(\mnhi,X)}

\def\ltp{\left ( \,}

\def\rtp{\, \right  ) }

\def\nhi{$N_{\rm HI}$}

\def \lndhall {$\log \rm{D/H} = -4.55 \pm 0.04$}
\def \obh {$\Omega_{b}h^{2}$}

\begin{document}

\title*{Baryons: What, When and Where?}
\author{Jason X. Prochaska and Jason Tumlinson}
\institute{Jason X. Prochaska \at
University of California Observatories -
Lick Observatory, University of California, Santa Cruz, CA 95064,
\email{xavier@ucolick.org}
\and Jason Tumlinson \at
Yale Center for Astronomy and Astrophysics, Yale University,
P. O. Box 208121, New Haven, CT 06520,
\email{jason.tumlinson@yale.edu}}

%
%
\maketitle

\abstract{
We review the current state of empirical
knowledge of the total budget of baryonic matter in the Universe
as observed since the epoch of reionization.
Our summary examines on three milestone redshifts since the
reionization of H in the IGM, z = 3, 1, and 0, with emphasis on the
endpoints.
We review the observational techniques used to discover and characterize the phases of baryons. In the spirit of the meeting,
the level is aimed at a diverse and non-expert audience
and additional attention is given to
describe how space missions expected to launch within the next decade will
impact this scientific field.
}

\section{Introduction}

Although baryons are believed to be a minor constituent
of the mass-energy budget of our universe, they have played
a dominant role in astronomy because they are the
only component that interacts directly and frequently with light.
Indeed, much of modern astrophysics focuses on the production
and destruction of heavenly bodies comprised of baryons.
Current observations of baryons extend from our Solar System to the
very early universe ($z>7$), i.e.\ spanning over 95$\%$ of the lifetime of our known
universe.  In terms of cosmology, the principal areas of baryonic research
include measuring their total mass density, identifying the various
elements and phases that comprise them, and resolving their
distribution throughout the universe.
In turn, astronomers are increasingly interested in probing
the baryonic processes that feed galaxy formation and the
feedback processes of galaxies that transform and return baryons
to the intergalactic medium (IGM).
Our mandate from the organizers was to review our
knowledge of baryons to a broad astronomical audience
from the epoch of reionization ($z\approx 6 - 10$) to the present day.
Granted limited time, we focus on
our empirical knowledge and limit the discussion of
theoretical inquiry.

In the past two decades, advances in telescopes and
instrumentation have significantly advanced our understanding
the distribution of baryonic matter in the cosmos.
Modern surveys of the local universe
yield an increasingly complete census of galaxies
and the large-scale structure in which they reside,
the distribution of galaxy clusters and mass estimates of
the hot gas within them,
and a view of the diffuse gas that lies between galaxies.
In this proceeding we review this work, placing particular
emphasis on techniques related to observing diffuse baryonic
phases.  Presently these techniques are most efficiently pursued
at $z \sim 3$ and $z \sim 0 - 0.5$, and observational
constraints on the properties and distribution of
baryons are most precise at these epochs.
It is somewhat embarrassing that there lies a nearly
10\,Gyr gap in our knowledge of the majority of baryons
between these two epochs.  Future missions need
to address this hole in addition to the primary uncertainties
remaining for the $z \sim 0$ and $z \sim 3$ epochs.

This proceeding is organized as follows: In $\S$~\ref{sec:primer}
we provide a basic introduction to the observational techniques
used to trace baryons throughout the universe.  The current best
estimates for the total baryonic matter density is reviewed in
$\S$~\ref{sec:omgb}.  We review the observational constraints on
the main phases of baryons at $z=3$ in $\S$~\ref{sec:z3}, briefly
comment on our general absence of knowledge at $z=1$ ($\S$~\ref{sec:z1}),
and then review the $z \sim 0$ missing baryons problem in $\S$~\ref{sec:z0}.
We finish with a brief review of planned, proposed, and desired mission
and telescopes that would significantly impact this science.

\section{A Primer on Tracing Baryons}
\label{sec:primer}

A complete baryon census poses a diverse set of observational
challenges to astronomy, where we most often study baryons by
observing the radiation that they emit, absorb or reflect.  This includes
stars, HII regions, cluster gas, planets, debris disks, etc.
With these observations,
one can assess the luminosity, temperature,
chemical composition, and size of the object under study.
In terms of
performing a complete census and analysis of baryons in the present and past
universe, however, this approach is currently limited by several
factors. First, there is the simple fact that more distant objects are
fainter.  Although recent advances in telescopes, instrumentation,
and search strategies now provide samples of galaxies at high redshift,
these samples are sparse and (at $z>1$) are comprised of only the
brightest objects. There is no {\em a priori} reason why the budget of
baryons would be dominated by the bright luminous objects, so these
samples do not necessarily account for a large fraction of the budget.
Second, it is generally difficult to estimate precisely the
mass of an emitting object.  With stars and galaxies, for
example, mass estimates generally rely on stellar evolution modeling which
is subject to many uncertainties (e.g.\ the initial mass function)
that are difficult to resolve even in the local universe.
Third, it turns out that only a small fraction of the universe's
baryons are in the collapsed, luminous objects that are most easily detected.
This point is especially true in the young universe but it even holds
in our modern universe. And finally, although diffuse gas emits line
and continuum radiation, it is very difficult to detect this radiation
even from gas in the nearby universe.

It is the last reason, in particular, that has driven observers
to absorption-line techniques for characterizing the mass density,
temperature, and distribution of the bulk of the baryons in the universe.
The absorption-line experiment is standard observational astronomy
in reverse: one observes a bright, background source (e.g.\ a quasar or
gamma-ray burst) with the aim to study the absence of light due to
absorption by intervening baryons. These basic analytic techniques
were developed primarily to study the interstellar medium (ISM) of our
Galaxy (e.g. \cite{spitzer78}) via UV spectroscopy of bright O and
B stars. Baryons with at least one electron will exhibit both resonant line
absorption (e.g.\ \lya) and continuum opacity (e.g.\ the Lyman limit
feature), primarily at UV and X-ray frequencies. The principal observable
of an absorption line is
its equivalent width $W_\lambda$, the fraction of light over a
spectral interval that is absorbed by the gas.  The principal physical
parameter is the column density $N$, the number of atoms per unit
area\footnote{Although we generally view stars and quasars as point
sources, they have finite area.  One may envision the concept of column
density like a core sample of the Earth, where spectra resolve
the location of gas in velocity not depth.} along the sightline.  This
is the number density equivalence of a surface density.
For weak absorption, the column density scales linearly with the equivalent
width.  For stronger (i.e.\ saturated) lines, $N$
depends weakly on $W_\lambda$ and may be difficult to estimate.
To assess these issues, it is highly desirable to obtain data
at spectral resolution and to measure directly the
optical depth profile of an absorption feature.  In this manner,
one can integrate the column density
provided accurate knowledge of the atomic data of the transition.

Table~\ref{tab:ions} presents the principal absorption line features
used to study H, He, and metals in diverse phases. Because most of these
spectral features have rest wavelengths at ultraviolet or higher energies,
one must utilize UV and X-ray telescopes to probe the past $\approx 7$\,Gyr
of the universe (i.e.\ $z<1$). These satellite missions have limited aperture
and (often) low-throughput instrumentation which requires very bright
background sources. Most of the gas, at least at $z>1$, is relatively cool
($T \sim 10^4$K) and the observed absorption lines have widths less
than $\approx 40 \mkms$. To spectrally resolve such features, one is driven
toward echelle spectrometers and generally the largest telescopes on Earth
(or in space). In these respects the $z\sim 0$ absorption-line experiment
is much more challenging than studies of the high $z$ universe, where
several of the key features are redshifted into optical pass-bands and
accessible to large ground-based telescopes. This has led, in  a few
ways, to a greater understanding of baryons in the $z \sim 3$ universe
than our recent past.

A tremendous advantage of the absorption-line experiment
is that it achieves extremely sensitive limits for studying
diffuse gas.  For example, the signal-to-noise and resolution easily
afforded by current telescopes and instrumentation allows
detection of the HI \lya\ transition to column densities
$\mnhi \sim 10^{12} \cm{-2}$.
This is ten {\it orders} of magnitude lower than the surface density
of a typical molecular cloud in a star-forming galaxy.  With
coverage of the full Lyman series, one can measure the
surface density of atomic hydrogen across these ten orders of magnitude
and probe the densest material to very diffuse environments.
By the same token,
one can study cold H$_2$ ($T<100$K) molecules through
Lyman/Werner absorption bands while also probing diffuse, hot
($T > 10^6$K) gas via metal-line transitions of O, N, and Ne.
One thereby traces structures in the universe with a wide dynamic
range of sizes: molecular clouds with diameters less than $100$\,pc in the
centers of galaxies to `voids' spanning tens of Mpc.

\begin{table}[t]
\caption{Key Baryonic Diagnostic Lines and Features}
\label{tab:ions}
\begin{tabular}{lccccccl}
\hline\noalign{\smallskip}
 Line & Phase & $T$ & $\lambda_{rest}$ & $\lambda_{z=1}$ & $\lambda_{z=3}$ & $\lambda_{z=9}$ \\
         && (K) &   (\AA) & (\AA)  &  (\AA) & ($\mu$m) \\
\noalign{\smallskip}\svhline\noalign{\smallskip}
Lyman-Werner&  Molecular gas      & 10--100       & $\sim$1000 & 2000 & 4000 & 1 \\
21cm        &  Atomic gas         & 100--1000     & 21cm & 0.7 Ghz & 0.4 Ghz & 140 MHz \\
\lya        &  Atomic+Ionized gas & 100--40000    & 1216 & 2400 & 4800 & 1.2 \\
H$\alpha$   & Ionized gas         & 10000--40000  & 6560 & 13000 & 26000 & 65000 \\
Lyman limit & Ionized gas         & 10000--40000  & 912  & 1800 & 3600 & 0.9 \\
HeII        & Ionized gas         & 10000--40000  & 304  & 450  & 912  & 0.2 \\
CIV         & Ionized Gas         & 20000--40000  & 1550 & 3000 & 6000 & 1.5 \\
OVI         & Warm/Hot Gas        & 20000--10$^6$ & 1030 & 2000 & 4000 & 1   \\
OVII,OVIII  & Hot Gas             & 10$^6$--10$^8$& 21.6,18.9   & 40   & 8    & 200 \\
NeVIII      & Hot Gas             & 10$^7$        & 775  & 1550 & 3100 & 7750 \\
\noalign{\smallskip}\hline\noalign{\smallskip}
\end{tabular}
\end{table}

There are several disadvantages, however, to absorption-line studies:
(1) the majority of key diagnostics have rest-frame UV or X-ray frequencies.
At low $z$, one requires a space mission with large
collecting area and efficient spectrometers to perform the
experiment.  This is expensive and often technologically infeasible;
(2) One requires a bright, background UV/X-ray source\footnote{Traditionally, researchers have
used quasars although increasingly transient gamma-ray burst
afterglows are analyzed (e.g. \cite{vel+04,cpb+05}).} subjecting
the final analysis to biases in the discovery and selection of
these sources;
(3) Bright UV/X-ray sources are rare and sparsely distributed
across the sky.  Therefore, multiplexing has limited value
and observational surveys are relatively expensive.  Furthermore,
with only a limited number of sightlines to probe the universe,
the densest (i.e.\ smallest) structure are at best sparsely sampled;
(4) The technique is limited to the most
distant luminous objects detected, currently $z \approx 6$;
(5) dust intrinsic to the gas under study may extinguish the
background source and preclude the analysis altogether.

\begin{figure}[b]
\sidecaption
\includegraphics[scale=0.4,angle=90]{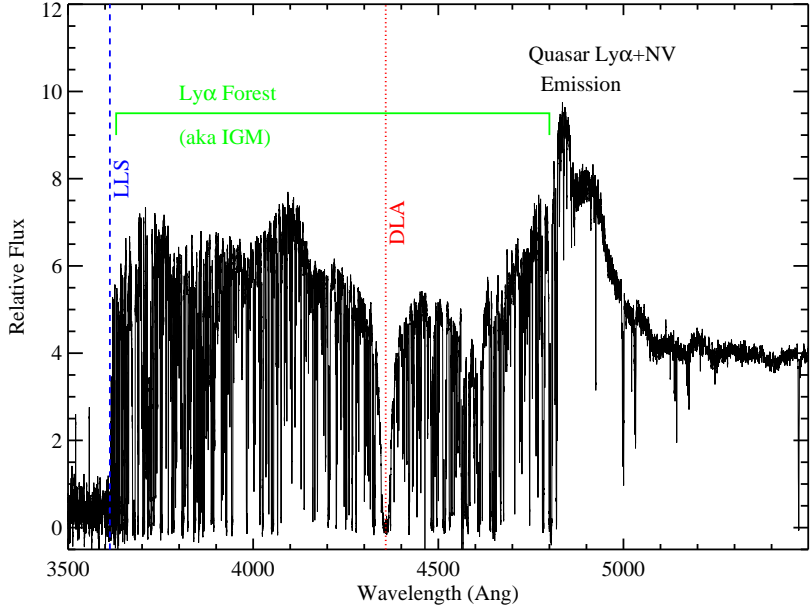}
\caption{Magellan/MIKE echelle spectrum of the $z_{QSO}=2.980$ quasar
J133941.95+054822.1.  The data reveal the broad \lya+NV emission
line of the quasar imposed on its otherwise nearly power-law spectrum.
Imprinted on this spectrum is a plethora of absorption features
due to gas foreground to the quasar.  In particular, one identifies
continuum opacity at $\lambda < 3600$\AA\ due to a Lyman limit
system at $z_{LLS}=2.97$ and a thick of absorption features related
to the \lya\ forest.  We also mark the damped \lya\ profile
at $\lambda \approx 4360$\AA\ of a foreground galaxy
($z_{DLA} = 2.59$).
}
\label{fig:qal}
\end{figure}

Several key aspects of absorption-line research are illustrated
in Figure~\ref{fig:qal} which presents the echelle spectrum of
quasar J133941.95+054822.1 acquired with the
Magellan/MIKE spectrometer \citep{bernstein03}.
This optical spectrum shows the rest-frame UV emission of the
$z_{QSO}=2.980$ quasar.  One notes the broad \lya+NV emission line
superimposed on a underlying, approximately power-law spectrum.
All of the absorption features, meanwhile, are associated with
gas foreground to the quasar.  The most prominent is the
continuous opacity at $\lambda < 3600$\AA\ which marks the
Lyman limit of a gas `cloud' at $z_{LLS} = 2.97$.  This Lyman limit
system (LLS) must have an HI column density $\mnhi > 10^{17} \cm{-2}$
to exhibit a Lyman limit opacity $\tau_{LL} > 1$.  The thicket
of narrow absorption lines at $\lambda = 3600$ to 4850\AA\ is
dominated by \lya\ absorption from gas with $z < z_{QSO}$.  These
lines are termed the \lya\ forest and they trace the
intergalactic medium (IGM) of the high $z$ universe.
Their HI column densities are generally less than
$10^{15}$ atoms per cm$^2$.
Within the \lya\ forest one also notes a very broad absorption
line at $\lambda \approx 4360$\AA\ whose observed equivalent width
is $W_\lambda \approx 50$\AA. This `redwood' of the \lya\ forest is a
damped \lya\ system (DLA); its very large HI column density,
$\mnhi > 2\sci{20} \cm{-2}$, allow the Lorentzian damping wings to
be resolved in the \lya\ transition. This feature marks the presence
of a foreground galaxy and has a larger HI column density than total of all
other lines in the \lya\ forest.
Finally, one observes a more sparse set of absorption features
at $\lambda > 4900$\AA\ which are metal-line transitions
from ions and atoms
of O$^0$, Si$^+$, C$^+$, C$^{+3}$, Fe$^+$, etc.
These lines enable studies of the metal abundance of our universe
\citep{schaye03,pgw+03} and also probe gas phases that
are difficult to examine with \lya\ alone.

\section{The Baryonic Mass Density}
\label{sec:omgb}

Our first problem is to assess the total budget of baryonic matter
in the Universe, preferably by independent measurement. The baryonic
mass density $\rho_b$, generally defined relative to the critical density
$\rho_c = 9.20\sci{-30} h_{70}^2 \; {\rm g \, cm^{-3}} $,
$\Omega_b \equiv \rho_b/\rho_c$ and $h_{70}$, the Hubble constant in units of
$70$ km s$^{-1}$ Mpc$^{-1}$, are fundamental cosmological parameters.
The first successful method for estimating $\Omega_b$
is to compare measurements of isotopes for the
light elements (H, D, $^3$He, $^4$He, Li) against
predictions from Big Bang Nucleosynthesis (BBN) theory.
The relative abundances of these isotopes are sensitive
to the entropy of the universe during BBN, i.e.\ the
ratio of photons to baryons.
At first, observers focused on He even though Li and
D are much more sensitive `baryometers'. This
was because He/H was inferred from nearby emission-line galaxies
where one could obtain exquisite
S/N observations.  It is now realized, however,
that the precision of the inferred He/H ratio is limited by
systematic uncertainties in the modeling of
HII regions and the precision of the atomic data \citep{plp07}.

\begin{figure}
\sidecaption
\includegraphics[scale=0.5]{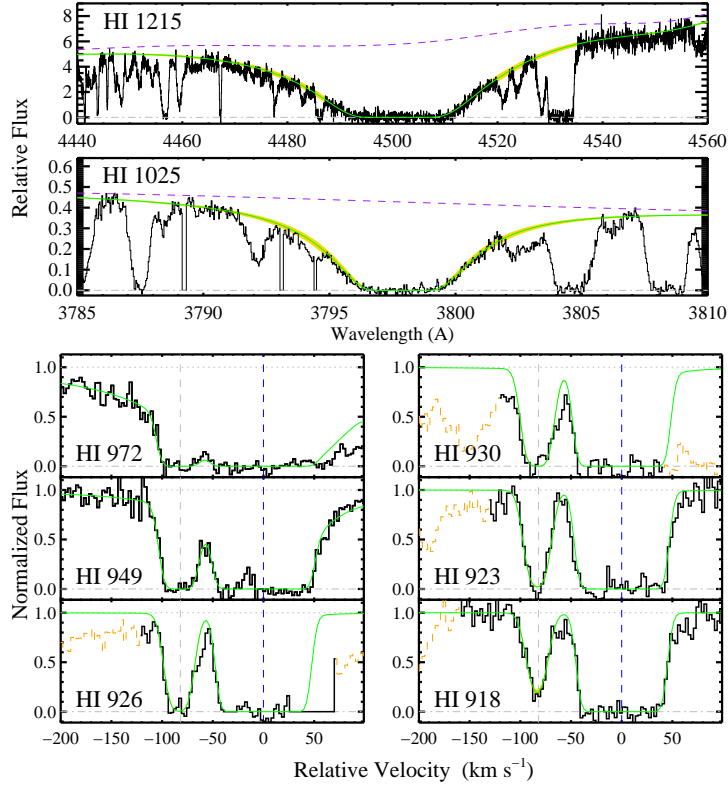}
\caption{H~I and D~I Lyman series absorption
in the $z = 2.70262$ DLA towards \dhobject\ \cite{obp+06}.
For the \lya\ and \lyb\ transitions,
the data is unnormalized and the dashed line traces the
estimate for the local continuum level.
The remaining Lyman series transitions are shown continuum normalized.
The solid green line shows the best single-component fit to the D~I and
H~I absorption. The estimate of the HI column
density comes from analysis of the damping wings
present in the Lyman $\alpha$--$\delta$ lines
whereas constraints on the DI column density
come from the unsaturated
D~I Lyman-11 transition ($\sim $918 \AA\ rest wavelength).
}
\label{fig:DH_HI}
\end{figure}

\begin{figure}
\sidecaption
\includegraphics[scale=0.45,angle=90]{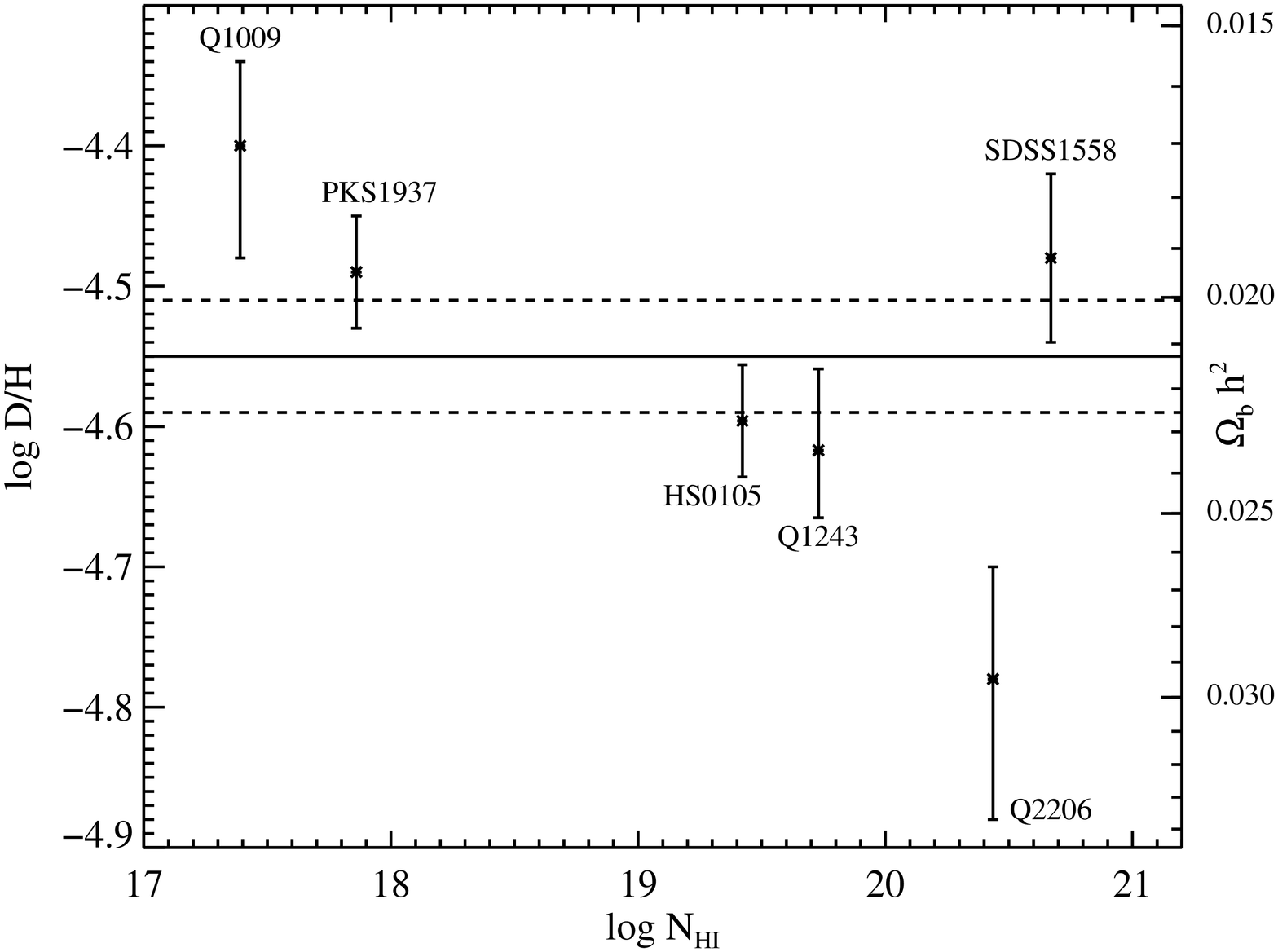}
\caption{Values of the D/H ratio vs. \nhi \cite{obp+06}.
The horizontal lines represent the weighted mean and
jackknife errors of the 6 measurements, \lndhall.
The right hand axis shows how the values of D/H translate
into values for \obh\ using BBN.
}
\label{fig:DH_all}
\end{figure}

The tightest constraints on $\Omega_b$ using BBN theory
come from direct measurements of the D/H ratio.  Ideally
this ratio is measured
in a metal-poor gas to minimize the likelihood that
D has been astrated by stars.  The first detections
of D, however, were from absorption-line studies in the metal-rich
Galactic ISM \citep{ry73}.
These observations establish a lower limit to the primordial
D/H value of $\approx 1\sci{-5}$ (e.g. \cite{moos02}).
In the past decade,
echelle spectrometers on 10m-class telescopes
have extended the experiment to high redshift \cite{bt98}.
Researchers have searched for the very rare `clouds'
which are metal-poor and therefore minimally astrated, have
large HI column density, and are sufficiently quiescent
that the higher order lines DI and HI
Lyman series are resolved (the velocity separation is 82\,\kms).
Figure~\ref{fig:DH_HI} shows a recent example \cite{obp+06}.
The first few transitions of the HI Lyman series
are broadened by the damping wings of their `natural'
Lorentzian profiles and a fit to the profile establishes
$\mnhi = 10^{20.67 \pm 0.05} \cm{-2}$.
The DI transitions are lost
within the cores of these strong absorption lines but higher
order lines show resolved and unsaturated
DI absorption.   Analysis of these lines provide
a precise estimation of the DI column density.

The D/H ratio is estimated
from the $\N{DI}/\N{HI}$ ratio under the reasonable assumption
that these isotopes have identical ion
configurations\footnote{ The gas is self-shielding and likely has a
high neutral fraction.} and differential depletion.
Allowing for these assumptions, the estimation of D/H is direct:
abundances measurements from absorption-line studies
are simply a matter of counting atoms.
Unlike most other methods to measure abundances,
the results are essentially independent of the
local physical conditions in the gas.
Figure~\ref{fig:DH_all} presents the `gold-standard' set of D/H
measurements from quasar absorption line surveys of high $z$, metal-poor
absorbers.  The observations are scattered about
log~D/H~$=-4.55$ with a boot-strap dispersion of 0.04\,dex.
Adopting standard BBN theory, this constrains
$\eta \equiv n_b/n_\gamma = 5.8 \pm 0.7 \sci{-10}$ where $n_b$
and $n_\gamma$ are the number densities of baryons and photons
respectively.   Direct measurements of the cosmic microwave
background give $n_\gamma$, and one derives a baryonic matter
density relative to the critical density $\rho_c$,
$\Omega_b = \rho_b / \rho_c = (0.043 \pm 0.003) h^{-2}_{70}$.
It is a triumph of the BBN theory and observational
effort that this value has been confirmed by observations of the
CMB power spectrum \citep{netterfield02,wmap05}.

Nevertheless, a few outstanding issues remain.  The dispersion
in D/H at high $z$ exceeds\footnote{Interestingly, the same
is true of the Galactic ISM measurements \citep{jenkins99}.  This
result is unlikely to be due to observational error and the
nature of the dispersion remains an open question
\citep{pth05,linsky06}.}
that predicted from the reported errors on the individual
measurements \citep{kts+03}.  Although this
most likely reflects over-optimistic error estimates by the
observers, it is worth further exploring models of inhomogeneous
BBN \citep{jedamzik04} and systematic error associated with the
astrophysics of H and D \citep{draine06}.  The other outstanding
issue concerns an inconsistency between the BBN-predicted Li abundance
based on the $\eta$ value from D/H and observation \citep{ssm84,bsd05}.
It is now generally expected that this observed Li underabundance
will be explained by non-primordial stellar astrophysics (e.g. \cite{korn06}),
but a comprehensive model of these processes
has not yet been developed.

\section{Baryons at $z \sim 3$}
\label{sec:z3}

This section reviews constraints on the
baryon census during the few Gyr following reionization,
starting with the densest
phases and proceeding to the least dense.  Perhaps not coincidentally (for hierarchical
cosmology), this organization also
proceeds roughly from the lowest cosmological mass density
to the largest.  Along the way, we highlight
the key uncertainties in these values and comment on
the impact of future missions.

\subsection{The Stellar Component}

The majority (but not all) of galaxy samples identified
at $z>2$ are discovered through spectral or photometric
signatures associated with ongoing star-formation.  These
include the Lyman break galaxies (see Shapley's presentation),
the \lya\ emitters (e.g. \cite{gfl+07}), sub-mm galaxies
\citep{cbs+05}, and gamma-ray burst host
galaxies (e.g. \cite{ldm+03}).
These populations dominate star formation activity during
the young universe.   It is likely that they also dominate the
stellar mass density in this epoch because `red and dead' galaxies
are very rare at this time (e.g. \cite{kvf+06}).
It is difficult, however, to estimate the stellar mass density
of these star-forming galaxies.
This is especially true when one is limited to optical
and near-IR photometry which corresponds to rest-frame UV and
blue light, i.e.\ photons from massive star formation that
provide little, if any, information on the low-mass end of the IMF.

To infer the stellar mass, one must adopt a mass to light ratio ($M_*/L$).
Standard practice is to model the stellar population that gives
rise to the photometry and spectroscopy of the galaxy.  This
modeling is sensitive
to the presumed star-formation history, reddening by
dust intrinsic to the galaxy, and uncertainties in the initial
mass function.   Altogether, these lead to large systematic
uncertainties in the stellar mass estimate.  These issues aside,
current surveys do not probe the fainter galaxies of what
appears to be a very steep luminosity function (e.g. \cite{rsp+07}).
The contribution by Dickinson provides a deeper
discussion of these issues.  Current estimates of the stellar
mass density at $z=3$ range from
$\Omega_* = 2$ to $4\sci{-7} \; h_{70} \mmsun \, {\rm Mpc^{-3}}$
with a systematic uncertainty of $\approx 50\%$
\cite{rlf+06,fsg+06}.
This corresponds to $0.005 \pm 0.002 \Omega_b h_{70}$.
The launch of new satellites with greater sensitivity in
near and mid-IR pass-bands (Herschel, JWST) will reduce
the uncertainties by observing the stellar component at
redder wavelengths and also by extending surveys to much
fainter levels.
Even with the large current uncertainty, however,
it is evident that stars
make a nearly negligible contribution to the total baryon budget
at these redshifts.

\subsection{Molecular Gas}

In the local universe, star formation is associated with
molecular gas in the form of molecular clouds.  Although the
link between molecular gas and star formation has not been
extensively established at $z \sim 3$, it is reasonable to
expect that the large SFR density observed
at this epoch implies a large
reservoir of molecular gas within high $z$ galaxies (e.g. \cite{hrt05}).
Indeed, molecular gas is detected in CO emission from a number of $z>2$
galaxies \citep{tnc+06,wbc+03} with facilities such as CARMA, IRAM, and CSO.
The observations focus on CO and other trace molecules because
the H$_2$ molecule has no dipole moment and therefore
very weak emission lines.
Unfortunately, current millimeter-wave observatories 
do not afford extensive surveys for CO gas in high $z$ galaxies.
The Large Millimeter
Telescope (LMT) and, ultimately, the Atacama Large Millimeter Array  (ALMA)
will drive this field wide open. These facilities will determine
the CO masses for a large and more representative sample of high $z$
galaxies and (for a subset) map out the spatial
distribution and kinematics of this gas.
Similar to the local universe, to infer the total molecular gas mass
from these CO measurements one must apply a CO to H$_2$ conversion
factor, the so-called `X-factor'.
The X-factor is rather poorly constrained even in the local universe
and nearly unconstrained at high $z$.  Therefore, a precise
estimate of the molecular component at high $z$ will remain
a challenging venture.

Although observations of H$_2$ in emission are possible 
at cosmological redshift\citep{oaa07}, these
measurements are not yet common enough to assess the total 
molecular gas budget at $z \sim 3$.
The H$_2$ molecule does exhibit
Lyman/Werner absorption bands in the rest-frame UV that are strong enough to
be detectable in many astrophysical environments with UV and optical spectrographs.
These ro-vibrational transitions have large oscillator
strengths and permit very sensitive searches for the
H$_2$ molecule, i.e.\ H$_2$ surface densities of less
than $10^{15}$ particles per cm$^2$.  To date \citep{gb97,petit00},
the search for H$_2$ has been targeted toward gas with
a known, large HI column density ($\mnhi > 10^{19} \cm{-2}$),
i.e.\ the Lyman limit and damped \lya\ systems.
This is a sensible starting point;  any sightline
intersecting molecular gas in a high $z$ galaxy would also
be likely to intersect the ambient ISM of that galaxy and
exhibit a large HI column density.  Although present
surveys are small, it is evident that the molecular fraction
in these HI absorbers is very low.  The frequency of
positive detections is $\approx 15\%$ \citep{nlp+08}
and the majority of these have molecular fractions
$f({\rm H_2}) < 10^{-3}$ \citep{ledoux03}.
Only a handful of these galaxies show molecular fractions
approaching the typical Galactic values of even diffuse molecular
clouds, $f({\rm H_2}) \sim 0.1$,
along sightlines with comparable HI column density \citep{cui05}.

In hindsight the paucity of H$_2$ detections,
especially ones with large $f({\rm H_2})$, is not
surprising.  The formation of H$_2$ clouds represents
a transition from the atomic phase, one that
leads to gas with much lower $T$ and much higher
density.  Consequently, cold molecular clouds phase exhibit
much lower cross-section than when the gas is atomic.
In turn, these clouds will have significantly lower probability
of being detected by absorption line surveys.  Analysis
of the distribution of molecular gas in the local
universe suggests that fewer than 1 in 1000 quasar sightlines
should pierce a cold, dense molecular cloud  \citep{zp06}.
In addition to this probalistic argument,
these clouds generally have very high dust-to-gas ratios
and correspondingly large extinction.  Therefore, even
in the rare cases that a quasar lies behind
the molecular cloud of a high $z$ galaxy, it may be reddened
and extinguished out of the optical, magnitude-limited surveys
that dominate quasar discovery.
It is somewhat curious, however, that GRB afterglow spectra
also show very low molecular fractions in gas that is believed
to lie very near star-forming regions \citep{tpc+07}.

In summary,
there are very limited current constraints on the
mass density and distribution of molecules in the young universe.
Furthermore,
future studies will focus primarily on tracers of H$_2$
and will also be challenged to precisely
assess this very important phase.
At present, our best estimate for the molecular mass density
may be inferred from the atomic and stellar phases.
Allowing for an order of magnitude uncertainty, we
can estimate $\Omega_{\rm H_2} = (0.001 - 0.1) \Omega_b$.
Therefore, we suggest that the molecular
phase contributes less than $10\%$ of the mass density
budget at $z \sim 3$.

\subsection{Neutral Hydrogen Gas}
\label{sec:neut3}

In the local universe, galaxies have roughly equal measures
of molecular and atomic hydrogen gas.  The latter phase
is primarily studied through 21-cm emission line surveys
using large radio telescopes.  Unfortunately, these radio telescopes
do not have enough collecting area to extend 21-cm emission-line
observations beyond $z \approx 0.2$ \citep{lcb+07}.  Although
one can trace the 21-cm line in absorption to much greater redshifts
(e.g. \cite{bw83,kc03}), detections are exceedingly
rare.  This is due in part to the paucity of bright background radio sources
and in part because galaxies have a relatively low
cross-section of cold, neutral gas.
For these reasons, neutral hydrogen gas at high $z$
is principally surveyed via \lya\ absorption. A total HI column density of
$\approx 10^{20} \cm{-2}$ is required for self-shielding to maintain 
neutral gas with a standard gas-to-dust ratio against
the extragalactic UV background and local UV sources
\citep{viegas95}. At these column densities, the damping wings
of the \lya\ transition are resolved even with moderate
spectral resolution (FWHM~$\approx 2$\AA; Figure~\ref{fig:qal}).
This spectral feature lends the name `damped \lya\ system' to
the absorbers with $\mnhi \ge 2\sci{20} \cm{-2}$.

\begin{figure}
\sidecaption
\includegraphics[scale=0.4,angle=90]{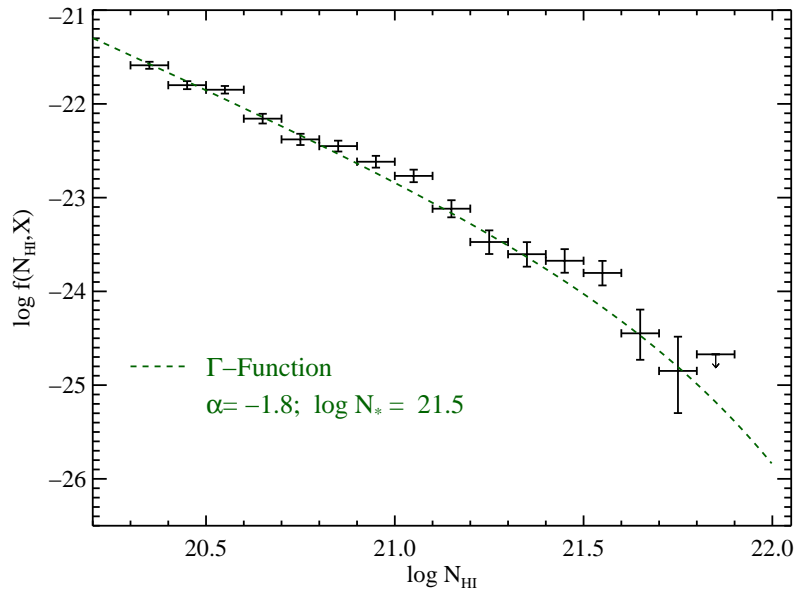}
\caption{The HI
frequency distribution \fnhi\ for the damped \lya\
systems identified in the SDSS-DR5 quasar database
(mean redshift $z=3.1$, \cite{phw05}).
Overplotted on the
discrete evaluation of \fnhi\ is the best fitting
$\Gamma$-function which is a reasonable (albeit
unphysically motivated) description of the observations.
}
\label{fig:fnDLA}
\end{figure}

Wolfe and collaborators pioneered damped \lya\ research beginning
with the Kast spectrometer on the 3m Shane telescope at Lick Observatory
\citep{wolfe86,wgp05}.
These optical observations cover redshifts $z>1.7$ which shifts Ly$\alpha$
above the atmospheric cutoff.
The most recent DLA surveys leverage the tremendous public database
of quasar spectroscopy from the Sloan Digital Sky Survey
(SDSS; \cite{sdssdr5}).  Figure~\ref{fig:fnDLA} presents the
frequency distribution \fnhi\ of \nhi\ values per $d\mnhi$ interval
per absorption pathlength $dX$ for 719 damped \lya\ systems
identified in the SDSS-DR5 database
(see http://www.ucolick.org/$\sim$xavier/SDSSDLA; \cite{phw05}).
This distribution function, akin to the luminosity function of
galaxies, is reasonably well described by a Gamma function
with a `faint-end slope' $\alpha = -2$ and a characteristic
column density $N_* \approx 10^{21.5} \cm{-2}$.
The first moment of \fnhi\ yields the mass density of neutral
gas $\Omega_{neut}(X) dX \equiv \frac{\mu m_H H_0}{c \rho_c}
  \intl_{N_{min}}^{\infty} \mnhi \mfnhi \, dX $
where $\mu$ is the mean molecular mass of the gas (taken to be 1.3),
$N_{min}$ is the column density which marks the transition from neutral
to ionized gas
(taken here as $10^{20.3} \cm{-2}$),
$H_0$ is Hubble's constant, and $\rho_c$ is the critical mass density.
The faint-end slope of \fnhi\ is logarithmically divergent, but the
break at $N_*$ gives a finite density.

Current estimates for \omg\ from the SDSS-DR5 database give
$\Omega_{neut} = 0.016 \pm 0.002 \Omega_b$.
There is a modest decline by a factor of $\approx 2$ from $z=4$
to $z=2$ where the latter value is (remarkably) in agreement
with the neutral gas mass density observed today ($\S$~\ref{sec:HIz0}).
One may speculate, therefore, that the neutral gas mass density has been
nearly invariant for the past $\approx 10$\,Gyr (but see \cite{rtn06}).
This mild evolution aside,
we conclude that the neutral, atomic gas phase at $z=3$
comprises only $\simeq 1-2$\% percent of the baryon budget at high $z$.
Cosmological simulations suggest that this gas comprises the
ISM of galaxies in formation, an assertion supported by measurements
of DLA clustering with bright star-forming galaxies \citep{bouche+05,cwg+06}
and the detection of heavy elements in all DLAs \citep{pgw+03}.
We can conclude that the sum of galactic baryonic
components is a minor fraction ($<10\%$) of the total baryonic mass density
at $z \sim 3$.
This is, of course, a natural consequence of hierarchical cosmology
(e.g. \cite{nwh+07}).

With the DLA samples provided by the SDSS survey, the statistical
error on \omg\ approaches $10\%$ in redshift bins $\Delta z = 0.5$.
At this level, systematic errors
become important, especially biases associated with
the selection criteria and magnitude limit of optically bright
quasars.  The bias which has received the most attention is
dust obscuration \citep{oh84,fall93}, i.e.\ dust within DLAs redden and extinguish
background quasars removing them from optical surveys.
The dust-to-gas ratio in the ISM of these high $z$ galaxies is small
\citep{pettini94,ml04,vpw08}, however,
and the predicted impact on \omg\ is only of the order of 10\%.
This expectation has been tested through DLA surveys toward
radio-selected quasars \citep{eyh+01,jwp+06}, which show similar
results, albeit with poorer statistical power.

Although we would like to extend
estimates of \omg\ to $z>5$, the experiment is challenged by the
paucity of bright quasars (although this may be overcome with GRB
afterglows) and because the collective, blended
opacity of gas outside of galaxies (i.e., the IGM) begins
to mimic damped \lya\ absorption.
By $z=6$, the mean opacity of the universe approaches unity at \lya\
\citep{wbf+03} and the experiment is entirely compromised.  Of course, current
expectation is that the universe is predominantly neutral not
long before $z=6$, i.e.\ $\Omega_{neut} = \Omega_b$.
These expectations and the evolution of neutral gas at $z>6$
will be tested through 21cm studies of the young universe
(see contributions by Furlanetto and Loeb for discussions of
this reionization epoch).

\subsection{Ionized Gas}
\label{sec:z3IGM}

\begin{figure}
\sidecaption
\includegraphics[scale=0.4,angle=90]{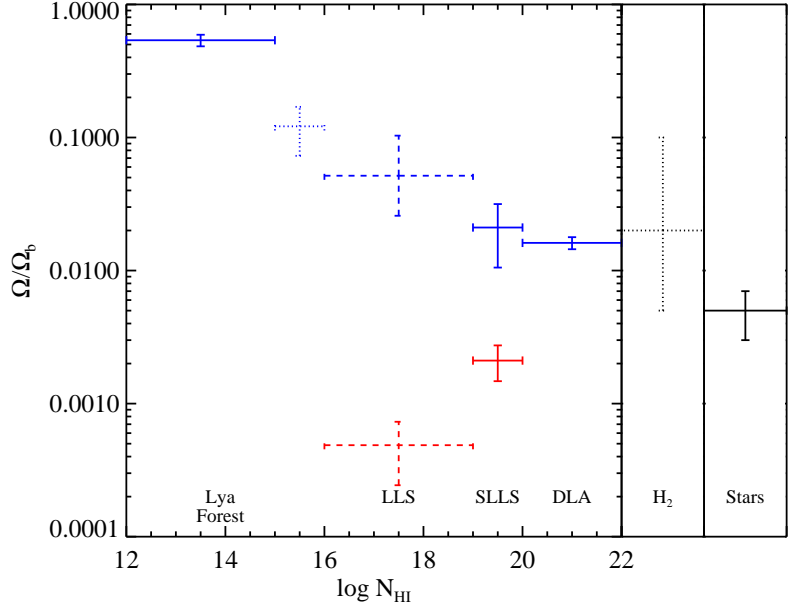}
\caption{Estimates of the baryonic mass density (relative to $\Omega_b$)
for various phases of baryons in the $z=3$ universe.
On the right side of the figure we show the stellar mass density
estimate from several recent works \citep{rlf+06,fsg+06}.
The uncertainty in this measurement is dominated by uncertainties
in stellar population modeling and the presumed IMF.
The phases corresponding to $\log \mnhi < 22$ refer to
absorption line systems traced by HI gas.
The molecular gas content is very crudely estimated to lie
near the stellar and atomic mass densities.
The neutral atomic phase is traced by the DLAs where one
estimates a mass density on the order of 1 to 2$\% \Omega_b$
\citep{phw05}.  Gas with $\mnhi < 10^{20} \cm{-2}$ is
predominantly ionized and one must make significant ionization
corrections to estimate the mass density from the observed
HI atoms \cite{opb+07}.  The red data points show the mass density
from HI atoms alone for the LLS whereas the blue
points represent estimates of the total mass density (see the
text for details).  The dashed symbols have weak empirical constraints
and the dotted symbols have very weak empirical constraints.
}
\label{fig:z3omg}
\end{figure}

As Figure~\ref{fig:z3omg} indicates, the mass densities of stars,
molecular gas, and neutral gas are unlikely to contribute significantly
to $\Omega_b$ at $z \sim 3$.  Unless one invokes an exotic form
of dense baryonic matter (e.g.\ compact objects), the remainder
of baryons in the early universe must lie in a diffuse component
outside of the ISM of galaxies \cite{petit93}.  
Indeed, the absence of a complete
Gunn-Peterson trough in $z < 6$ quasars demonstrates that the majority
of baryons are highly ionized at these redshifts\citep{fck06}.
Because it is impossible to directly trace H$^+$ for the vast majority of
the mass in the diffuse IGM, we must probe this phase through the remaining trace
amounts of HI gas or, in principle, HeI and
HeII  Unfortunately, the resonance-line
transitions of He have wavelengths that preclude easy detection
(Table~\ref{tab:ions}; though the Gunn-Peterson effect has
been observed in HeII at $z > 2.1$, see \cite{kso01,stg04}).
Therefore, the majority of research has
focused on the frequency distribution of HI column
densities for gas associated with the IGM (a.k.a.\ the \lya\ forest).

\begin{figure}
\sidecaption
\includegraphics[scale=0.4,angle=90]{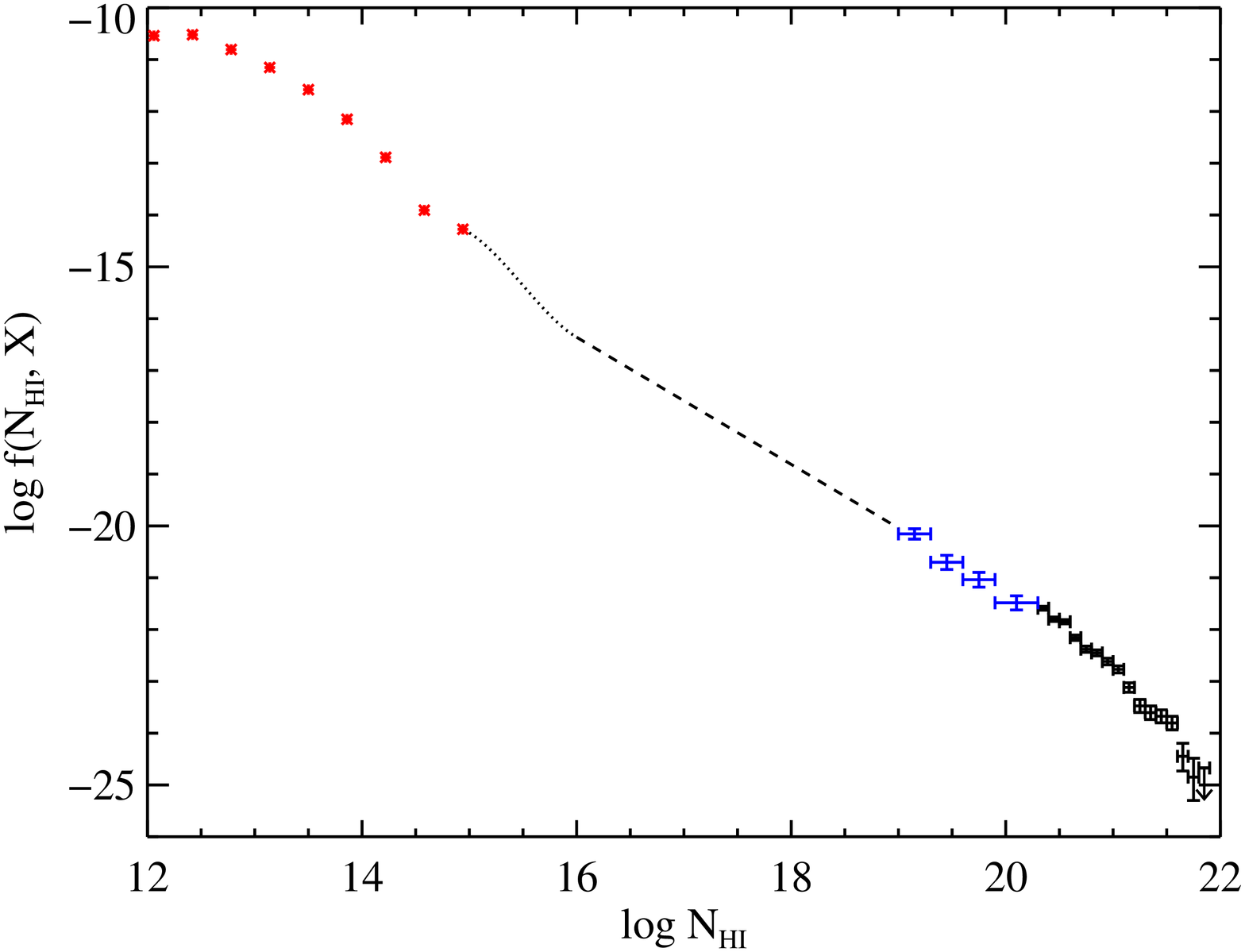}
\caption{The \fnhi\ distribution function of ionized
$\mnhi < 10^{20} \cm{-2}$ and neutral $\mnhi > 10^{20} \cm{-2}$
at $z=3$.  The data points at $\mnhi \ge 2 \sci{20} \cm{-2}$
correspond to the damped \lya\ systems and are drawn
from the SDSS-DR5 analysis of \cite{phw05}.  The data
at $19 < \log(\mnhi/\cm{-2}) < 20.3$ correspond to the
super Lyman limit systems (SLLS) and were drawn from the
analysis of \cite{opb+07}.  The dashed line traces
their best guess at the frequency distribution for
the Lyman limit systems with $16 < \log(\mnhi/\cm{-2}) < 19$.
For the \lya\ forest ($\mnhi < 10^{15} \cm{-2}$) we adopt
the results from \cite{kt97} normalized to our assumed
cosmology \citep{wmap03}.
Finally, the dotted line is a spline fit to the \lya\ forest
data and the functional form for the LLS given by \cite{opb+07}.
Although the turnover in \fnhi\ at $\mnhi < 10^{13.5} \cm{-2}$
may be partially due to incompleteness, it is likely a real
effect (see also \cite{hkc+95,kcc+02}).
}
\label{fig:fnall}
\end{figure}

The modern description for the \lya\ forest is an undulating density
fields with minor, but important peculiar velocities and
temperature variations.  In this paradigm, it may be more
fruitful to assess distribution functions of the smoothly varying optical
depth of the IGM \cite{msb+06,brs07,kbv+07,faucher08}
and compare these against theoretical models \citep{mco+96,bhv+05}.
Nevertheless, the classical approach of
discretizing the \lya\ forest into individual
absorption lines does yield valuable insight into the distribution
and mass density of baryons in the intergalactic medium.
This analysis is generally summarized through the distribution
function of HI column densities, \fnhi.
Figure~\ref{fig:fnall} summarizes our current knowledge
of the \fnhi\ distribution for both the ionized and neutral gas.
Quasar absorption line analysis to date has provided an assessment of \fnhi\
from $\mnhi = 10^{12}$ to $10^{22} \cm{-2}$ with an important
gap spanning from $\mnhi \approx 10^{15-19} \cm{-2}$.  At
column densities larger than $10^{22} \cm{-2}$,
the systems are too rare to be
measured with existing quasar samples. Detections below
$10^{12} \cm{-2}$ are limited by S/N and by the difficulty of
distinguishing absorption from single `clouds'
from unresolved blends of weaker and stronger lines (e.g. \cite{kt97}), while at
$10^{11} \cm{-2}$, the clouds are unphysically large for typical IGM
density and ionization conditions.

Let us focus first on the systems exhibiting
the largest \nhi\ values, the super Lyman limit systems (SLLS;
also referred to as sub-DLAs).  These absorbers are
defined to have $10^{19} \cm{-2} < \mnhi < 10^{20.3} \cm{-2}$.
Similar to the DLAs, one can measure the HI column densities of SLLS
through fits to the damping wings of the \lya\ profile although
higher resolution spectroscopy is required \citep{peroux_slls03,opb+07}.
Current surveys indicate that \fnhi\ flattens below
$\mnhi \approx 10^{20} \cm{-2}$ with $d\log f/ d\log N \approx -1.4$
\citep{opb+07}. If the SLLS are predominantly neutral, then they
would contribute only a few percent to $\Omega_{neut}$ and a negligible fraction of $\Omega_b$
(Figure~\ref{fig:z3omg}). Most of these absorbers, however, are
partially ionized
with mean ionization fraction $<x> \approx 0.9$
\cite{pro99,pdd+07}.
In Figure~\ref{fig:z3omg} we display the baryonic
mass density of the SLLS inferred from the HI frequency distribution
and assuming an \nhi-weighted, average ionization fraction $<x> = 0.9$.
We infer that the SLLS contribute on the order of $2\%$ of the
baryonic mass density at $z=3$.
The majority of the baryons must reside in yet more diffuse and
highly ionized gas.

At column densities $\mnhi = 10^{16}$ to $10^{19} \cm{-2}$
(the Lyman limit systems; LLS),
the majority of the Lyman series lines are saturated and the damping
wings of \lya\ are too weak to measure.
For these reasons, it is difficult to estimate \nhi\ values in
this interval.  In principle, one can precisely measure \nhi\ values for
the partial Lyman limit absorbers with $\mnhi < 10^{17} \cm{-2}$ whose opacity
at the Lyman limit is $\tau_{912} \approx 1$.  To date, however,
only small samples from heterogeneous surveys exist in this range of
$\mnhi$ (e.g. \cite{petit93}).
Because the Lyman limit feature is easily identified, one more
easily derives an integral constraint on \fnhi\ by surveying the incidence
of LLS in quasar spectra $\ell(X) = \int \mfnhi d\mnhi$
(e.g. \cite{storrie94}).
This integral constraint includes gas with $\mnhi > 10^{19} \cm{-2}$,
but is dominated by absorbers with low \nhi\ values.
Current estimates indicate $\ell(X)=7.5$ at $z=3$ \citep{peroux_dla03}.
Analysis of the SDSS database will greatly improve this estimate for
$z>3.2$ (O'Meara et al., in prep).  Furthermore, complete analysis
of the \lya\ series for LLS will yield tighter constraints on
\fnhi\ within the interval $\mnhi = 10^{17}$ to $10^{19} \cm{-2}$
(Prochter et al., in prep.). \cite{opb+07} combined the current observational constraints
to predict \fnhi\ across the LLS regime.  These are presented as
the dashed curve in Figure~\ref{fig:fnall} and
the uncertainty in this estimation is $\approx 50\%$.

Not only is \fnhi\ very uncertain for the LLS, so too is the
ionization state of this gas.  For a few systems, analysis of metal-line transitions
have confirmed the gas is predominantly ionized
with ionization fractions $x > 0.95$ (e.g. \cite{pb99}).
The samples are too small, however, to reveal the distribution or
trend of ionization fraction $x$ with \nhi.  Unfortunately, there is little
insight from theory.
Previous work with cosmological simulations
has severely underestimated
the incidence of Lyman limit systems (e.g. \cite{gkh+01}) and
it is likely that a careful treatment of radiative transfer
is required to assess these systems that are near the transition
from neutral to ionized \cite{kg07}.
We can set a very conservative lower limit
to their contribution by taking the \fnhi\ function of
O'Meara et al.\ and assume $x = 0$, $\Omega_{LLS} > 0.0004 \Omega_b$.
Ionization corrections will undoubtedly increase this limit by one or
more orders of magnitude.  Even in the case of a 100 times increase,
it is very unlikely that LLS dominate the baryonic budget at $z \sim 3$.
Figure~\ref{fig:z3omg} presents an estimate of the baryonic mass density
of the LLS where we have adopted the HI frequency distribution
presented in Figure~\ref{fig:fnall} and estimated the ionization correction
from simple Cloudy calculations assuming an ionization parameter $\log U = -2.5$,
inferred from metal-line analysis of several LLS \citep{pb99}.
This implies a mass density estimate of
$\approx 0.05 \Omega_b$.

For column densities $\mnhi \approx 10^{12}$ to $10^{14.2} \cm{-2}$,
where \lyb\ and/or \lya\ are optically thin, the frequency
distribution is reasonably well characterized by a power-law
$\mfnhi \propto \mnhi^{-1.5}$ (e.g. \cite{kt97,kcc+02}).
Observers do report, however, a departure from this power-law
at $\mnhi < 10^{13} \cm{-2}$ which is likely not simply a
consequence of incompleteness in the analysis \cite{kt97}.
Although the \lya\ forest absorbers dominate the spectrum of
any quasar by number, the total number of HI atoms
that they contribute is negligible ($\approx 4\sci{-6} \Omega_b$).
From cosmological simulations of the high-$z$ universe, we expect that
the \lya\ forest arises in undulating density fields of highly ionized gas filling
large regions ($\approx$Mpc) with modest overdensity $\delta \rho / \rho \sim 3$.
These same simulations argue that the majority of the baryonic
mass density resides in the \lya\ forest, with a contribution
possibly exceeding 0.9\,$\Omega_b$ (e.g. \cite{mco+96,rauch97,bma+99}).
Empirically, the gas is observed to only exhibit high-ion states
of elements like C, Si, and O which demonstrate a high degree
of ionization.

Because this gas is optically thin, if one can estimate
its volume density and the intensity of the ionizing radiation
field it is relatively straightforward to estimate its ionization fraction
and thereby the total baryonic mass density of the \lya\ forest.
\cite{schaye01_lya} has presented a simple but intuitive prescription
for modeling the \lya\ forest as gravitationally
bound gas clouds whose size is of order the Jeans length.
Within this prescription, which matches scaling
laws derived from cosmological simulations, the ionization
fraction $x$ of the gas can be related to the
HI column density:

\begin{equation}
x = 1 \; - \; 4.6 \sci{-6} \;
\ltp\frac{\mnhi}{2\sci{13} \cm{-2}}\rtp^{2/3} \; T_4^{-0.6} \; \Gamma_{12}^{-1/3}
\label{eqn:x}
\end{equation}
where $T_4$ is the gas temperature in units of 10$^4$ K
and $\Gamma_{12}$ is the photoionization rate in units
of 10$^{-12} \; {\rm s^{-1}}$.

Photoionization should maintain
the gas at a temperature near $2 \sci{4}$K, as predicted
by the cosmological simulations.  Within this prescription,
therefore, the only major uncertainty is the ionization rate of
the extragalactic UV background (EUVB) radiation field, $\Gamma_{EUVB}$.
At $z=3$, the EUVB will have contributions from both quasars
and star-forming galaxies.  The contribution of the former
is directly measured from the luminosity function
of $z=3$ quasars.   The contribution of galaxies, however, is
far more uncertain.  Although recent studies offer relatively
well constrained UV luminosity functions (e.g. \cite{rsp+07}),
these are measured at $\lambda \approx 1500$\AA\ in the rest frame, and
so must be extrapolated to estimate the flux of ionizing
photons.   Furthermore, one must adopt an
escape fraction $f_{esc}$ to estimate the flux emitted by
these star-forming galaxies.
Current empirical constraints are extremely limited
\citep{ssp+06,stc+07,cpg07}; the estimates range from
$f_{esc} = 0$ to $\approx 10\%$.  Because the comoving
number density of star-forming galaxies greatly exceeds that
of quasars, it is possible that galaxies contribute as much
or more of the ionizing flux to the EUVB. Thus this ionizing background
is uncertain at high redshift, with a corresponding uncertainty
in the ionization correction for diffuse ionized gas.

The most conservative approach toward estimating the baryonic
mass density of the \lya\ forest is to assume
$\Gamma_{EUVB} = \Gamma_{QSO}$.
\cite{hrh07} provide an
estimate of $\Gamma_{QSO} = 4.6 \sci{-13} \; {\rm s^{-1}}$ at $z = 3$
with an $\approx 30\%$ uncertainty.
Convolving Equation~\ref{eqn:x} (assuming $\Gamma_{12} = 0.46, T_4 = 2$)
with \fnhi, we infer a baryonic mass density
$\Omega_{\rm Ly\alpha} = 0.65 \Omega_b$.  Again, this should
be considered a lower limit to $\Omega_{\rm Ly\alpha}$ subject
to the uncertainties of Equation~\ref{eqn:x}.
More detailed comparisons of the opacity of the \lya\ forest
with cosmological simulations reach a similar conclusion:
the EUVB ionization rate exceeds that from quasars alone and
$\Omega_{\rm Ly\alpha} = 0.95 \Omega_b$ (e.g. \cite{mco+96,jnt+05,meiksin08}).
From a purely empirical standpoint, however, we must allow that
as many as $20\%$ of the baryons at $z=3$ are still unaccounted for.
It is very unlikely that the difference lies in photoionized gas with
$\mnhi < 10^{12} \cm{-2}$ because the mass contribution of the
\lya\ forest peaks near $\mnhi = 10^{13.5} \cm{-2}$ \citep{schaye01_lya}.
Instead, one would have to invoke yet another phase of gas,
e.g.\ hot diffuse baryons.

Ongoing surveys will fill in the current gap in \fnhi\
for $\mnhi = 10^{19} \cm{-2}$ down to at least
$\mnhi = 10^{16.5} \cm{-2}$.  This gas, which likely
is the interface between galaxies and the IGM, is very
unlikely to contribute significantly to the baryonic
mass density.  It may be critical, however, to the
metal mass density at $z \sim 3$ \citep{poh+06,bouche06II}.
Furthermore, the partial Lyman limits ($\mnhi \sim 10^{17} \cm{-2}$)
dominate the opacity of the universe to HI
ionizing radiation and therefore set the ``attenuation length''\citep{fgs98}.
This quantity is necessary to convert the observed luminosity
function of ionizing radiation sources into an estimate of $\Gamma_{EUVB}$
\citep{hm96,mw03}.
These observations may also provide insight into the
processes of reionization at $z>6$ and the escape
fraction of star-forming galaxies.

\subsection{Hot Diffuse Baryons}

In the previous section, we demonstrated that the photoionized IGM
may contain all the baryons in the $z \sim 3$ universe
that are not associated with dense gas and stars.
We also noted, however, that this conclusion hinges on the
relatively uncertain intensity of the EUVB radiation field.
As a consequence even
as much as $30\% \Omega_b$ may be unaccounted for in the IGM.
One possible reservoir in this category is a hot $(T>10^5$K) diffuse phase.
Indeed, there are several processes already active at $z \sim 3$ that
can produce a reservoir of hot and diffuse gas.
The hard, intense radiation field from quasars and
other energetic AGN processes (e.g.\ radio jets) may heat and ionize
their surroundings out to
several hundred kpc (e.g. \cite{cmb+08}).  Similarly,
the supernovae that follow star formation
may also drive a hot diffuse medium into the galactic halo and perhaps
out into the surrounding IGM (e.g. \cite{ahs+01,kr07}).
Outflows of ionized gas are observed in the spectra of quasars
and star-forming galaxies at these redshifts.
On the other hand, gas accreting onto galactic halos is expected to shock
heat to the virial temperature ($T>10^6$K), especially in halos with
$M > 10^{11} M_\odot$ \citep{db06}.
Altogether these processes may heat a large mass of gas in
and around high $z$ galaxies.

Presently, there are very weak empirical
constraints on the mass density and distribution of a hot,
diffuse medium at $z \sim 3$.  Detecting this gas in emission is not technically
feasible and there are very few observable
absorption-line diagnostics.
difficult to trace, even with absorption-line techniques.
The most promising approach uses the
OVI doublet to probe highly ionized gas associated with
the IGM (e.g. \cite{simcoe02,bergeron05}) and
galaxies \cite{simcoe06,fpl+07}.
These studies suggest a hotter component ($T >\sim 10^5$K)
contributes on the order of
a few percent $\Omega_b$ at $z \approx 2$.

The mass density of an even hotter phase ($\simeq 10^{6-7}$ K), meanwhile, is
unconstrained by observation.
Ideally we would assess this component with
soft X-ray absorption and/or emission but this will require a
major new facility such as Con-X.
We will take up these issues again in Section $\S$~\ref{sec:whim} on the $z \sim 0$
universe,
where the warm/hot component is predicted to be substantial.

\begin{table}
\caption{Empirical Summary of Baryons in the Universe}
\begin{tabular}{cccccc}
\hline\noalign{\smallskip}
Phase & Temperature & \multicolumn{2}{c}{$\Omega(z=3)$}
   & \multicolumn{2}{c}{$\Omega(z=0)$} \\
      &  (K)        & Location & Estimate$^a$   & Location & Estimate$^a$ \\
\noalign{\smallskip}\svhline\noalign{\smallskip}
Stars         & --     & Galaxies & $0.005 \pm 0.002$ & Galaxies   &  $0.06 \pm 0.03$ \\
Molecular Gas & $10^2$ & Galaxies & $>0.001$          & Galaxies   &  $0.0029 \pm 0.0015$ \\
Neutral Gas   & $10^3$ & Galaxies & $0.016 \pm 0.002$ & Galaxies   &  $0.011 \pm 0.001$ \\
Ionized Gas   & $10^4$ & IGM      & $>0.80$           & IGM        &  $0.17^{+0.2}_{-0.05}$ \\
Warm/Hot Gas  & $10^6$ & Galaxies?& $>0.01$           & Filaments? &  ??  \\
Hot Gas       & $10^7$ & ??       & ??                & Clusters   &  $0.027 \pm 0.009$  \\
\noalign{\smallskip}\hline\noalign{\smallskip}
\end{tabular}
\\
$^a$ Relative to an assumed $\Omega_b$ value of 0.043.
\end{table}

\section{Baryons in the $z\sim 1$ Universe}
\label{sec:z1}

In contrast to the success that astronomers have had in
probing baryons at $z \sim 3$, the knowledge of the distribution
of baryons at $z \sim 1$ is quite limited.
This is the result of several factors.
One aspect is psychological:  astronomers have pushed
harder to study the universe at earlier times.
This emphasis is shifting, however, with
advances in IR spectroscopy (where the rest-frame optical diagnostics
fall for $z \sim 1$), the impact of the Spitzer mission
(e.g.\ Dickinson's contribution), and the generation of large
spectroscopic samples of $z \sim 1$ galaxies (GDDS, DEEP, VIRMOS).
The other aspect is technical.
As with the $z \sim 3$ universe, it is
difficult or impossible with current facilities
to build large samples of clusters to assess the host phase, or to survey the
molecular phase with CO,
or to measure HI gas with 21-cm techniques.
Furthermore, the majority of key absorption-line
diagnostics (e.g.\ \lya, CIV; Table~\ref{tab:ions}) fall below the atmospheric
cutoff and UV spectroscopy on space-based telescopes
is necessary and expensive.  Key emission-line diagnostics
(H$\alpha$, [OIII]), meanwhile, shift into the near-IR making
it more difficult to pursue large galaxy surveys.

Surprisingly, the prospects are also limited for major advances in exploring
baryons at $z \sim 1$ during the next decade.
The notable positive examples are (i)
Sunyaev-Zeldovich experiments, which should establish the mass function
of $z \sim 1$ clusters,
(ii) multi-object, near-IR spectrometers which will improve
detections of star-forming galaxies and the luminosity function
of early-type galaxies and
(iii) the construction of mm telescopes
(LMT, ALMA) that will enable searches for molecular gas.
These advances will allow astronomers
to characterize the dense, baryonic component
at $z \sim 1$.
Similar to $z \sim 3$ and $z \sim 0$, however, we expect that these
components will be minor constituents.  It is a shame, therefore,
that there is currently no funded path to empirically constrain
the properties of diffuse gas, neutral or ionized.
Upcoming `pathfinders' for the proposed Square Kilometer Array (SKA)
will extend 21-cm observations beyond the local universe but
are unlikely to reach $z=1$.  This science awaits an experiment
on the scale of SKA.  The classical absorption-line experiments
must access hundreds of faint QSOs at $z \simeq 1 - 2$ and
so require a UV space mission probably on the scale of JWST
\cite{bsp}.
At present, however, there is no development funding path
toward such a mission.  In this
respect, our ignorance on the distribution and characteristics
of the majority of baryons in the $z \sim 1$ universe may
extend beyond the next decade.

\section{Baryons in the $z \sim 0$ Universe}
\label{sec:z0}

In this section, we follow closely the footsteps of
\cite{fhp98}, who presented a comprehensive census
of baryons in the local universe (see \cite{fukugita04} for an update).
With this exercise in accounting, the authors stressed a startling problem:
roughly 50\%\ of the baryons in the local universe are missing!
They further suggested that these missing baryons are likely
hidden in a warm/hot ($T \approx 10^5$ to $10^7$K) diffuse medium
that precluded easy detection.
These claims were soon supported
by cosmological simulations that predict a warm/hot intergalactic
medium (WHIM) comprising 30 to 60\%\ of today's baryons
\cite{co99,daveetal01}.  Motivated by these semi-empirical and
theoretical studies, the
search for missing baryons gained great attention and
continues today.  We review the current observational
constraints on the distribution and phases of baryons in
the local universe and
stress paths toward future progress.
As in previous sections, we roughly order the discussion
from dense phases to the diffuse.

\subsection{Stars}

Large-area imaging and spectroscopic
surveys (e.g., 2MASS, SDSS, 2dF) have precisely measured
the luminosity function of galaxies in the local universe
to faint levels \cite{cnb+01,blantonetal03}.  The
light emitted by galaxies in optical and IR bands, therefore,
is well characterized.  Establishing the mass budget
of baryons in these stars, however, requires translating the luminosity
function into a mass function\citep{sp99}.  The typical procedure is to
estimate a stellar mass-to-light ratio ($M/L$)
for each galaxy based on stellar population modeling of
the colors of the integrated light.
These estimates are sensitive to the metallicity, star-formation
history, and (most importantly) the initial mass function (IMF)
of star formation.
Because the stellar mass of a galaxy is dominated by lower
mass stars even at early times, the analysis is most accurately performed
in red passbands.

The SDSS and 2dF galaxy surveys have led the way in this line
of research; SDSS possesses the advantage of $z$-band imaging
($\lambda \approx 9000$\AA) which
reduces the uncertainty in the $M/L$ estimate. \cite{khw+03} analyzed
the mass-to-light ratio for a large sample of SDSS galaxies, for which
they report a luminosity function-weighted ratio $<M/L_z> = 1.5$ with an
uncertainty of $\approx 20\%$ and assuming a
Kroupa IMF. Taking the \cite{khw+03} estimation of $M/L$ with the SDSS
luminosity function \cite{blantonetal03},
\cite{fukugita04} estimates a stellar mass density $\Omega_*(z=0) = 0.0025$,
which includes an estimate of stellar remnants.
Assuming a 50\%\ uncertainty,
we find $\Omega_*(z=0) = (0.06 \pm 0.03) \Omega_b$.

Further progress in estimating the stellar contribution to $\Omega _b$
is not limited by neither statistical error nor depth of the luminosity function, but rather by systematic
uncertainty, especially in the IMF. \cite{khw+03} did not vary this
assumption when estimating the error budget.
Because sub-solar mass stars contribute most of
the mass in stellar systems, uncertainties in the IMF
can easily dominate the error budget.  A conservative estimate of this
uncertainty is $\approx 50\%$, especially if we allow for a Salpeter IMF.

The subject of stellar masses and their sensitivity to IMF is worthy of
a review article in its own right, and so it is well beyond
the scope of our mandate. Here we only pause to note three
recent developments. First, the uncertainty in {\em total} stellar masses is probably
only a factor of 1.5 - 2, even though the error may be larger for individual cases.
Second, this topic is receiving more attention in stellar population
synthesis models (e.g. Maraston, Bruzual and Charlot), which can now include IMF
uncertainties in their budgets of systematic error. Finally, there have been recent,
speculative suggestions that the IMF at high redshift may depart significantly from the Galactic form. At
early times, generally hotter environments within galaxies and an elevated CMB
may raise the typical stellar mass \cite{tum07,vdok08,dave08}.
Though much theoretical work and experimental
testing remains to be done in this area, it is possible that these ideas will
ultimately lead to a more self-consistent picture of the evolution of stellar mass
with redshift, with a consequent improvement in our understanding of the baryon budget.
This is one of the key science goals of the JWST and ALMA facilities. These uncertainties
aside, it is evident that stars are a sub-dominant baryonic component
in our current universe. As an editorial aside, we note the irony that even the baryonic component
that is easiest to see - the luminous stars - is difficult to weigh. This uncomfortable
fact should be kept in mind later as we attempt to assess the unseen,
hot diffuse phases of the IGM.

\subsection{Molecular Gas}

With the exception of the Galaxy \cite{MW_H2}, the Magellanic Clouds \cite{tsr+02},
and some nearby starburst galaxies \cite{hsh04}, there are very few direct detections
of H$_2$ in the local universe. Commonly these detections are of absorption in the FUV Lyman-Werner
bands, and so are accessible only in optically thin, low-dust ($A_V < 1$) environments. H$_2$
can also be detected in UV fluorescence lines from interstellar gas, but these are intrinsically very faint. Neither of these
$H_2$ reservoirs are likely to contribute much to the total cosmic mass budget of molecular gas
(judging from the Milky Way). Because it is a homonuclear molecule with no dipole moment,
H$_2$ has only very weak quadrupole emission even if the density is high. The usual
technique, then, is to track molecular gas with trace optically-thin species and infer the
total molecular content through scaling factors calibrated against our Galaxy and
others in the local universe. The most common indirect tracer is CO and its conversion factor
is termed the ``X-factor''. In contrast with HI gas (see below), `blind' surveys
for CO have not yet been performed. Instead, researchers have estimated the molecular mass
density by correlating CO mass with another galaxy observable (e.g.\ HI mass,
infrared flux density) that has a well defined, volume-limited distribution function.
The convolution of the two functions gives an estimate of the CO mass function
\cite{ys91,kyy03,zp06}.  Finally, this is converted to the H$_2$ mass function with
the X-factor which is generally assumed to be independent of any
galaxy properties.  This latter assumption is known to be invalid for faint metal-poor
galaxies and luminous starbursts, but variations are believed to be small for
the $L \approx L_*$ galaxies that likely dominate the molecular mass density
\cite{dy90}.

Taking $X \equiv \N{H_2}/I(CO) = 3\sci{20} \cm{-2} (\rm K \, km \, s)^{-1}$
\cite{ys91}, \cite{kyy03} report $\mohtwo = \rho_{\rm H_2} =
2.2 \sci{7} h_{70} M_\odot \, {\rm Mpc^{-3}}$ for an
IR sample of galaxies and \cite{zp06} report $\mohtwo = 1.2 \sci{7} h_{70}
M_\odot \, {\rm Mpc^{-3}}$ for the optically selected BIMA sample.
The factor of two difference in these estimates
reflects the systematic error associated with sample selection.
We adopt the mean value of the two estimates and
assume a 30\%\ uncertainty:
$\mohtwo = 0.0029 \pm 0.0009 \Omega_b$.
The next major advance for surveying molecular gas in the local
universe awaits a wide-field, deep, and blind survey of CO gas
with depth and area comparable to current 21-cm surveys.
This is a difficult technical challenge and we know of no such survey
that is currently planned.

\subsection{Neutral Hydrogen Gas}
\label{sec:HIz0}

Over the past decade, radio astronomers have performed
blind surveys for 21-cm emission to derive the HI mass
function of galaxies (e.g. \cite{zbs+97,rs02,zms+05}).
The modern version is the HIPASS
survey \cite{hipass}, an all Southern-sky survey for HI
with the Parkes radio telescope.  \cite{zms+05} present
the HI mass function from HIPASS and derive an integrated
mass density $\Omega_{\rm HI} = 3.8 \sci{-4} h_{70}^{-1}$.
Including He, this implies a neutral gas
mass density $\momg = (0.011 \pm 0.001) \Omega_b$.
Although some uncertainty remains regarding the shape
of the low-column (``faint-end'') slope for the HI
mass function, the total mass density is
dominated by $L \approx L_*$ galaxies.  Another important
result of these surveys is the demonstration that very little HI mass,
if any, associated with star-free systems.
Ultimately, the ongoing ALFALFA survey \cite{alfalfa1} will surpass
the HIPASS sample in sensitivity and
provide the definitive measure of the
HI mass function at $z \sim 0$.  In the future,
facilities like the EVLA or Square Kilometer Array will be able to
push such observations to $z \sim 1$ or beyond.

\begin{figure}[b]
\sidecaption
\includegraphics[scale=0.4,angle=90]{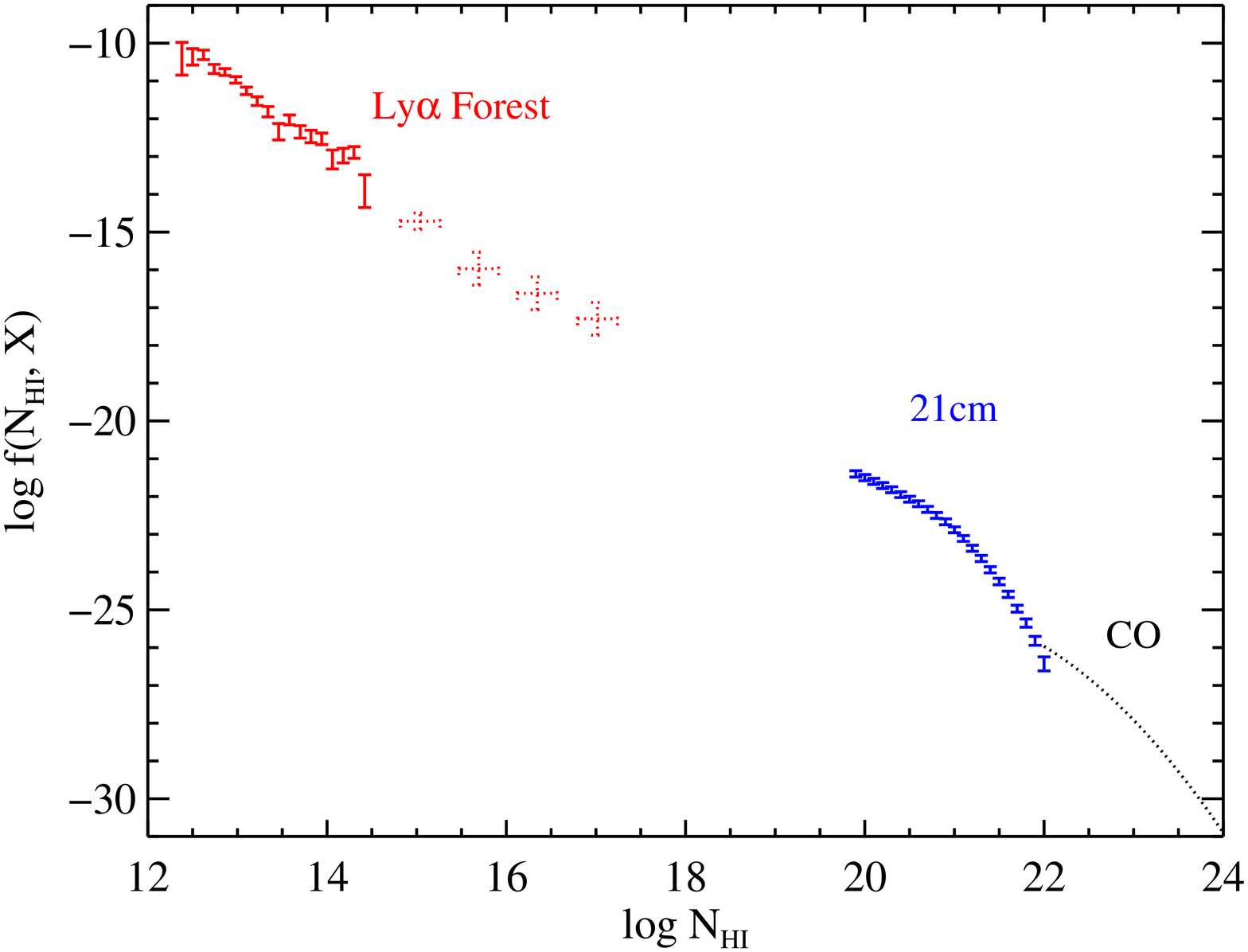}
\caption{The \fnhi\ distribution function of ionized
$\mnhi < 10^{20} \cm{-2}$, neutral $\mnhi = 10^{20-22} \cm{-2}$,
and molecular $\mnhi > 10^{22} \cm{-2}$ gas
at $z \sim 0$. The data at $\mnhi < 10^{17} \cm{-2}$ which correspond
to HST/STIS observations of the low redshift \lya\ forest are taken from
\cite{pss04}, the distribution function neutral, atomic hydrogen
gas is taken from 21cm WSRT observations \cite{zvb+05}, and the
distribution function of molecular gas is estimated from BIMA CO
observations \cite{zp06}.
}
\label{fig:z0fnall}
\end{figure}

Interferometric observations can map the HI
surface density profiles of low-redshift galaxies and thereby
infer the HI frequency distribution at $\mnhi > 10^{20} \cm{-2}$
\cite{rws03,zvb+05}.  This distribution function is
illustrated in Figure~\ref{fig:z0fnall}.  Remarkably, it has
nearly identical shape to the \fnhi\ function derived for
$z \sim 3$ galaxies from damped \lya\ observations
(Figure~\ref{fig:fnall}).  Both of these \fnhi\ functions follow a roughly
$\mnhi^{-2}$ power-law at lower column densities and exhibit a break
at $\log \mnhi \approx 21.7$.  This suggests that HI gas
is distributed in a similar manner within galaxies at $z \sim 0$
and $z \sim 3$.  Perhaps even more remarkable is the fact that
the integrated mass density is also very comparable.
The \omg\ value at $z \sim 0$ is at most three times
smaller  than the
largest \omg\ values derived from the $z \sim 3$ damped
\lya\ systems (0.0045 vs.\ 0.0014)
and actually coincides with the \omg\
value at $z=2.2$ \cite{phw05}.
One might infer that \omg\ has been
constant for the past $\approx 10$\,Gyr, which suggests
that all of the gas accreted into the neutral phase
within galaxies is converted into stars
(e.g.\ \cite{kkw+05}).

\subsection{Intracluster Medium (ICM)}

The majority of galaxies in clusters are early-type with little or no neutral
hydrogen gas. It may have been thought at first,
therefore, that the baryonic content of clusters
was almost entirely stars. X-ray observations, however, have revealed that clusters
are filled with a hot plasma that extends to at least
their virial radii ($\sim 1$ Mpc).
This intracluster medium (ICM) contains the baryons
that have collapsed within the cluster structure and have been shock-heated to the
virial temperature ($T > 10^7$K) of the cluster.
Analysis of the X-ray surface brightness indicate that
the ICM has an electron density profile
$n_e(r) = n_0 [1 + (r/r_c)^2]^{3\beta/2}$ with $\beta \approx 0.6$
and $n_0$ ranging from $10^{-1}$ to $10^{-2}$ at $r_c$ of a few hundred kpc
(e.g.\ \cite{rb02}).
In fact the ICM mass significantly exceeds the stellar mass inferred for
the galaxies within the cluster.
From a cosmological standpoint, therefore, the ICM may
represent a major baryonic component of our current universe.

Detailed analysis of the X-ray observations show that the
gas fraction, $f_{gas}$, or the ratio of the gas mass to the dynamical mass,
is nearly constant with cluster mass, $f_{gas} = 0.11 \pm 0.01$ \cite{ars+08}.
Therefore, one can estimate the
total baryonic mass density of the ICM in all clusters simply by taking
the product of $f_{gas}$ with the cosmological mass density of clusters.
A recent estimation of the latter comes from an
X-ray selected sample of clusters with overlapping SDSS observations
\cite{rdn08}. Fitting these data to the \cite{jfw+01} dark-matter halo
mass function for a WMAP5 cosmology \cite{wmap05}, and
integrating from high mass down to $M_{min} = 10^{14} \mmsun$, we
derive a mass density of clusters, $\Omega_{cluster} = 0.010$.
Adopting $f_{gas} = 0.11$, the ICM
has a baryonic mass density $\Omega_{ICM} = 0.027 \Omega_b$.
Uncertainties in this estimate are dominated by the uncertainty
in $\Omega_{cluster}$, which are in turn driven by uncertainty in the
power spectrum normalization specified by $\sigma_8$. These errors translate
into a roughly 30\% uncertainty in $\Omega _{cluster}$.
This diffuse, hot component represents a non-negligible
mass density in our modern universe.  It raises
the possibility that a similar medium associated with collapsed
structures having $M<M_{min}$ (i.e.\ groups and isolated galaxies)
could contribute significantly to the mass density of baryons at $z \sim 0$
\cite{mulchaey96,fhp98}.

\subsection{Photoionized Gas}

Summing the mass densities for the gas and stars of the
preceding sub-sections, we estimate a total mass density of only
$\approx 0.10 \Omega_b$.  Similar to the high-$z$ universe,
therefore, we conclude that the majority of baryons at $z \sim 0$
lie in diffuse gas outside of virialized structures.  It is reasonable
to speculate first that the photoionized
component, i.e.\  the intergalactic medium, is also the major baryonic
reservoir at $z \sim 0$.
As with the $z \sim 3$ universe, this component is best traced by
\lya\ absorption using UV absorption-line spectroscopy.
Even a visual inspection of low-$z$
quasar spectra shows
that the incidence of \lya\ absorption $\ell(X)$ is quite low relative
to the $z \sim 3$ universe.  This sharp decline in $\ell(X)$
is a natural consequence, however, of the expanding universe.
Quantitatively, the decline in $\ell(X)$ is actually less than predicted
from expansion alone \cite{wjl+98} because of a coincident decrease
in the EUVB intensity which implies a higher
neutral fraction and correspondingly higher incidence of HI
absorption \cite{dhk+99}.

The HI column density distribution function
of the $z\sim 0$ IGM has been characterized
by several groups using UV spectrometers onboard the Hubble Space
Telescope \cite{dt01,pss04}.  In Figure~\ref{fig:z0fnall}, we
show the Penton et al.\ \fnhi\ results for a sample with mean
redshift $z=0.03$. These results assume a Doppler parameter
$\blya = 25 \mkms$ for all absorption lines identified in their survey (\cite{ds08} have
obtained $\blya = 27 \mkms$ for their larger sample of 650 Ly$\alpha$ absorbers).
The results are insensitive to this assumption for low column densities
($\mnhi < 10^{14} \cm{-2}$) because the \lya\ profiles are unsaturated but
measurements at higher \nhi\ are sensitive
to $\blya$ and have much greater uncertainty.
Similar to the $z \sim 3$ universe, the \fnhi\ function
is sufficiently steep that one expects the mass density is dominated
by clouds with $\mnhi < 10^{15} \cm{-2}$.
\cite{ds05} have fitted the distribution function at low \nhi\
with a power-law function
$\mfnhi = A \mnhi^\beta$ and report
$A = 10^{10.3 \pm 1.0}$ and $\beta = -1.73 \pm 0.04$ for their recent sample
with $\langle z \rangle = 0.14$, and there is
significant degeneracy between the two parameters.

We may estimate the IGM baryonic mass density at $z \sim 0$ using
the formalism described in $\S$~\ref{sec:z3IGM} following \cite{schaye01}.
In addition to \fnhi, this analysis requires an
estimate of the HI photoionization rate $\Gamma$
and the gas temperature $\tlya$.  The dependences of the mass density
on these quantities are weak ($\olya \propto \Gamma^{1/3} T^{3/5}$)
but so too are the empirical constraints.
Estimates of $\Gamma$ range from 0.3 to $3 \sci{-13} \; {\rm s^{-1}}$
\cite{srg99,sbd+00,wvv+01} and we adopt
$\Gamma = 10^{-13} \; {\rm s^{-1}}$ based on calculations of the
composite quasar and galaxy background estimated by Haardt \& Madau (CUBA; in prep).
The characteristic temperature
of the IGM is constrained empirically only through the observed
distribution of line widths, $\blya$.
Several studies report a characteristic line width at $z \sim 0$
of $\blya \approx 25 \mkms$ which implies $\tlya < 38,000$K
\cite{dt01,pss02,pjt+08}.  This is an upper limit because
it presumes the line widths are thermally and not turbulently, broadened.
Indeed, \cite{dt01} have argued from analysis of numerical simulations
that the observed $\blya$ distribution is in fact dominated by non-thermal
motions.  These same simulations suggest a density-temperature relation
$\tlya \approx 5000 (\rho/\bar \rho)^{0.6}$K (the so-called ``IGM equation
of state'').

Combining $\Gamma = 10^{-13} \; {\rm s^{-1}}$ and
the \cite{dt01} $\rho-T$ ``equation-of-state'' with the \cite{pss04} estimation
for \fnhi\ and the \cite{schaye01} formalism, we estimate
the baryonic mass density of the $z \sim 0$ \lya\ forest
($\mnhi = 10^{12.5-14.5} \cm{-2}$) to be
$\olya = 0.0075 = 0.17 \Omega_b$. From their sample of 650 Ly$\alpha$ absorbers
at $\langle z \rangle = 0.14$, \cite{ds08} obtain $\olya = 0.0131 \pm 0.0017 = 0.29 \Omega_b$.
This range is still several times smaller than the mass density estimate for the IGM
at $z \sim 3$ (Figure~\ref{fig:z3omg}).  The immediate implication
is that the IGM may not be the majority reservoir of baryons at $z \sim 0$.
Indeed, the missing baryons problem hinges on this result.
It is crucial, therefore, to critically assess the uncertainty
in this estimate, in particular whether current observational
constraints allow for a significantly larger value.

First, consider the photoionization rate $\Gamma$ which is
constrained to no better than a factor of a few (at any redshift).
As noted above, $\olya$ is largely
insensitive to the photoionization rate ($\olya \propto \Gamma^{1/3}$).
Furthermore, our adopted value lies toward the upper end of current
estimations which leads us to conclude that a more precise measure of
$\Gamma$ will not markedly increase the baryonic mass density
inferred for the IGM at $z \sim 0$.

Another consideration is the temperature of the IGM, although again
the mass density is not especially sensitive to this quantity.
Even if the \lya\ forest were uniformly at $\tlya = 2\sci{4}$K,
we would derive only a 50$\%$ higher baryonic mass density.
It is evident that modifications to $\Gamma$ and $\tlya$ will
not make a qualitative difference in $\olya$, though these effects in combination
could increase it by a factor of 2.
Turning to the HI frequency distribution,
\cite{pss04} and \cite{dt01} emphasize that \fnhi\
shows no break from a power-law down
to the sensitivity limit of the HST/STIS observations,
$N_{lim} = 10^{12.5} \cm{-2}$.
It is reasonable, therefore, to extend the distribution function
to lower column densities.  Taking $N_{lim} = 10^{11.5} \cm{-2}$
with the power-law exponent fixed at $\beta = -1.65$, leads to a
$2\times$ increase in $\olya$.
Finally, both cosmological simulations and some observational analysis
suggest a steeper power-law
($\beta \approx -2$) \cite{dt01,lsr+07,pjt+08} which,
in turn, may lead to an enhancement in $\olya$.
If we allow for a combination of these effects, it is
possible to recover $\olya$ values exceeding $0.5 \Omega_b$.  For example,
the following parameter set -- $\Gamma = 10^{-13} \; {\rm s^{-1}}$,
$\tlya = 2\sci{4}$K, $\beta = -1.8$, $N_{lim} = 10^{11.3} \cm{-2}$ --
yields $\olya = 0.6 \Omega_b$.

On empirical grounds alone, then, the current observations do not
rule out a scenario where the photoionized
\lya\ forest contains most of the baryons not locked in denser phases, which
we estimate at $\simeq 0.1 \Omega _b$.
This would require, however, a warmer IGM than predicted
by numerical simulations and a \fnhi\ function that remains
steep to $\mnhi \approx 10^{11} \cm{-2}$.  The latter point
may be tested with new observations of the
IGM using the planned HST/COS instrument, especially if
it achieves S/N~$> 50$ per pixel.
Given that the uncertainties must be pressed in concert to their limits
for the budget to be complete,
it is also reasonable to conclude that the IGM
does not contain all of the missing baryons at $z \sim 0$. In the next section
we will take up the question of what could fill this gap, assuming it exists.

\begin{figure}
\sidecaption
\includegraphics[scale=0.4,angle=90]{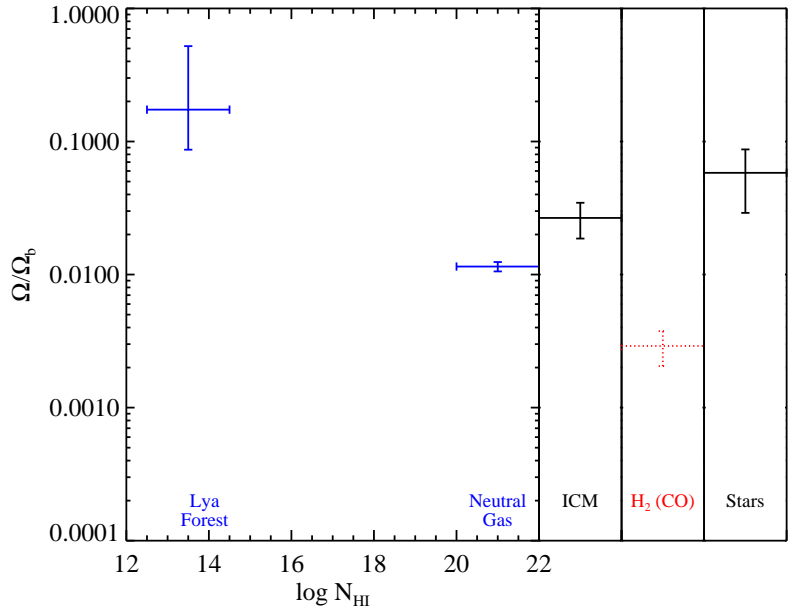}
\caption{Estimates of the baryonic mass density (relative to $\Omega_b$)
for various phases of baryons in the $z \sim 0$ universe.
On the left side of the figure we show the mass densities
estimated from the photoionized \lya\ forest and the neutral
gas in low $z$ galaxies.
One notes that the sum of the central values for the various
components is significantly less than unity.  This suggests
that a significant mass component is missing from this census,
e.g.\ the warm/hot intergalactic medium (WHIM).
Even adopting the maximum values for each component, one
may require an additional baryonic phase to contribute at $z=0$.
}
\label{fig:z0omg}
\end{figure}

\subsection{The WHIM}
\label{sec:whim}

In Figure~\ref{fig:z0omg}, we present the $z \sim 0$ baryonic mass density
estimates for the phases described in the previous sub-section.
A simple summation of the central value estimate of each
component\footnote{Although there lies a gap
in our knowledge or gas with $\mnhi \approx 10^{17} \cm{-2}$, it is unlikely
that this gas contributes significantly to $\Omega_b$.} yields a total
$\Omega_{stars} + \Omega_{H_2} + \momg + \oicm + \olya = 0.27 \Omega_b$.
Therefore, having followed the footsteps of FHP98, we reach a similar
conclusion: current observations have not revealed roughly 70\%\ of
the baryons in the $z \sim 0$ universe.
The observational techniques described in the previous section
are sensitive to baryons in luminous matter, hot ($T>10^7$K) and diffuse gas,
molecular and atomic neutral gas, and diffuse photoionized gas.
From a practical standpoint, this leaves one obvious phase: warm/hot
($10^5 \; {\rm K} < T < 10^7 \; {\rm K})$, diffuse gas.

Indeed FHP98 identified this phase as the likely reservoir for
the remainder of baryons and suggested it would be associated with
galaxy groups and galactic halos (see also \cite{mulchaey96}).
Cosmological simulations of the low redshift universe, meanwhile,
argue that a warm/hot phase exists yet external to galaxies, i.e.\
within the intergalactic medium \cite{co99,daveetal01}.
This warm/hot intergalactic medium (WHIM), believed to be
heated by shocks during large-scale gravitational collapse,
is predicted by the simulations to comprise roughly 50\% of
baryons in the present universe.
Bregman \cite{bregman07} has recently reviewed the observational and
theoretical evidence for the WHIM at $z \sim 0$.  We refer the
reader to that manuscript for a broader discussion.  We
highlight here only a few points with emphasis on absorption-line
observations.

The WHIM gas, as conceived in numerical simulations or as gas reservoirs
of galaxy groups, has too low density and temperature to be detected
in emission by current X-ray instrumentation and
even the next generation of X-ray facilities would be unlikely to
detect the most diffuse gas.  The neutral fraction,
meanwhile, is far too low to permit detections via 21cm observations.
For these reasons, searches for WHIM gas have essentially been limited to
absorption-line techniques.  The current approaches include (1)
a search for the wisps of HI gas associated with the WHIM and
(2) surveys for ions of O (and Ne) which trace warm/hot gas.

The first (and most popular) approach
has been to survey intergalactic O$^{+5}$ along quasar sightlines.
This is primarily driven by observational efficiency;
oxygen exhibits a strong OVI doublet at ultraviolet wavelengths
($1031,1037$\AA) that can be surveyed with high resolution, moderate
signal-to-noise observations.  To date, these have been acquired by spectrometers
on the HST and FUSE satellites.
In collisional ionization equilibrium (CIE; \cite{sd93,gs07}),
the O$^{+5}$ ion is abundant in gas with $T = 10^5$ to $10^6$K
and therefore may trace the cooler WHIM.
A hard radiation field, however, may also produce substantial
O$^{+5}$ ions in a diffuse gas.  Therefore, the detection of OVI is
not definitive proof of WHIM gas.
Tripp and collaborators were the first to demonstrate
that intergalactic OVI may be detected along quasar
sightlines \cite{trippetal00,ts00}.
Surveys along tens of sightlines have now been comprised
with HST/STIS and FUSE observations \citep{ds05,tc08a,ds08,tripp08}
which reveal a line density $\ell(z)$ of absorbers $\ell(z) \approx 10$ per
unit redshift path down to $W_{1031} \ge 50$m\AA\ and
$\ell(z) \approx 40$ down to $W_{1031} \ge 10$m\AA.

\begin{figure}
\includegraphics[scale=0.4,angle=0]{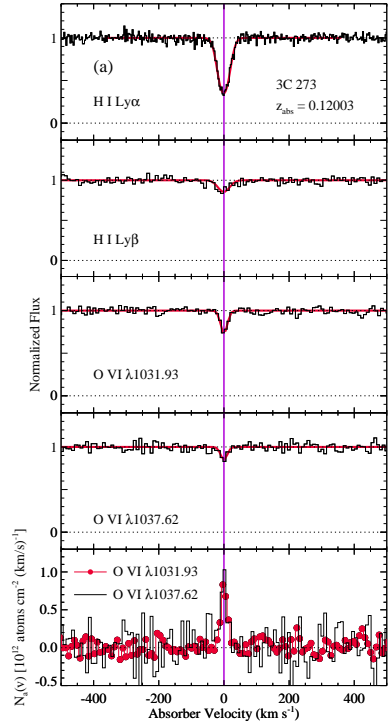}
\caption{Plot of the \lya, \lyb, and OVI line-profiles for an OVI
absorber at $z=0.12$ toward 3C273 (taken from \cite{tripp08}).
Note the high similarity between the OVI and HI profiles.  This
indicates the gas is not thermally broadened and requires a gas
temperature $T < 10^5$K.  In turn, this implies the material
is predominantly photoionized, not collisionally ionized and is
therefore unlikely to correspond to warm/hot intergalactic gas.
Roughly $40$\% of OVI absorbers share these properties.
It remains an open question as to whether the majority of observed
OVI gas corresponds to the WHIM.
}
\label{fig:ovi}
\end{figure}

Intriguingly this incidence and the observed equivalent width
distribution are in broad agreement with the predictions from
$z \sim 0$ cosmological simulations \cite{cf06}.
At the least, this should be considered circumstantial evidence
for WHIM gas. On the other hand, at least 30\% of OVI systems showing
OVI absorption also exhibit coincident CIII absorption
\cite{pks0405_uv,dsr+06,ds08,tc08b}.  The O$^{+5}$ and C$^{++}$ ions cannot coexist
in CIE; their coincident detection either indicates a multi-phase
medium or a photoionized gas.
Tripp et al.\ \cite{tripp08} have also emphasized that $\approx 40\%$ of
OVI absorbers show HI absorption with nearly identical line profile
(Figure~\ref{fig:ovi}, see also \cite{tc08b}). In their analysis, \citet{ds08} have
placed this fraction at $< 10\%$, perhaps owing to their choice to not
divide the absorbers into velocity subcomponents. Thus there is
agreement that at least some, if not all, of these absorbers have
temperatures of $T<10^5$K, i.e.\ photoionized material\footnote{Although
one could invoke undetected, broad \lya\ absorption in these cases,
this would not predict the observed identical line-profiles for OVI
and \lya\ absorption.}. In other cases, the OVI
absorption is offset in velocity from the strongest HI absorption and
there is little diagnostic constraint on the gas temperature.  In only
a few cases (e.g.\ the $z=0.1212$ absorber toward HS1821+643; \cite{trippetal01})
does one detect broad \lya\ absorption aligned with the OVI absorption
that clearly favors collisionally ionized gas with $T>10^5$K.

In support of the view
that the OVI absorbers are predominantly hot and collisionally ionized, \citet{ds08}
offer the arguments that (1) the column-density and linewidth distributions of the two
high ions OVI and NV differ significantly from those of commonly photoionized
species, such as CIII/IV and HI, and (2) the observed systems imply clouds with
an unrealistically wide range of ionization parameters in close proximity
to yield OVI and lower ions. Of the known OVI systems, the majority have not been confirmed
by multiple investigators as truly arising from collisional ionization, so
there is presently no consensus that OVI in fact measures the $T>10^5$ WHIM.
The upshot of these ongoing controversies is that while OVI can be a
tracer of WHIM, its origins are ambiguous and further work, including detailed cross-checking
by the interested groups, is needed to assess
the fraction of OVI absorbers that are truly hot and collisionally ionized,
and from there to calculate their contributions to the baryon budget.

It may in fact be premature to identify this OVI-traced gas with
the intergalactic medium.   Although the incidence of OVI
absorbers is too high for all of them to be associated with the
halos of $L \approx L^*$ galaxies,
there are sufficient numbers of dwarf galaxies to locate
the gas within their halos or ``zones of metal enrichment''
if these extend to $\approx 150 - 250$\,kpc \cite{tf05,ds08,stockeetal06}.
Indeed, searches for galaxies linked with OVI absorbers
have indicated a range of associations from individual
galactic halos to galaxy groups to intergalactic gas
\cite{tsg+05,pwc+06,tripp06,stockeetal06,cpc+08}.
We await a larger statistical sample and detailed comparison
to numerical simulations to resolve this issue \cite{gcf+08}.
The upcoming Cosmic Origins Spectrograph on HST will make a
large and important contribution to solving this problem.

Setting aside the controversial aspects of OVI, one may speculate on the
mass density of the OVI-bearing gas $\oovi$ using similar techniques
as for the \lya\ forest.  Because OVI is simply a tracer of the
material, one must assume both the metallicity and an ionization
correction for the gas.  Current estimations give
$\oovi \approx 0.1 \Omega_b$, assuming that all
the OVI detections arise in collisionally ionized gas \citep{ds08}.  Therefore, even
in the case that the OVI-bearing gas traces only WHIM material,
it is evident that it comprises a relatively small fraction of
the baryon census.

Ideally, one would assess the mass
density and spatial distribution of WHIM gas through observations of
hydrogen which dominates the mass.
This is, however, a difficult observational challenge.
Assuming CIE, a WHIM `absorber' with (an optimistic) total hydrogen
column of $N_H = 10^{20} \cm{-2}$ and $T=10^6$K will have an
HI column density of $\mnhi = 10^{13.4} \cm{-2}$ and
Doppler parameter $b_{HI} = 129 \mkms (T/10^6 \rm K)^{1/2}$.  This
gives a peak optical depth $\tau_0 = 1.5 \times 10^{-2}
(N \lambda_{\rm Ly \alpha} f_{\rm Ly\alpha}) / b_{HI} = 0.25$ and
a total equivalent width $W_{\rm Ly\alpha} = 137$m\AA.
The detection of such a line-profile requires at
least moderate resolution UV spectra at relatively high S/N
and also a precise knowledge of the intrinsic continuum
of the background source.

Detections of line profiles
consistent with broad \lya\ absorption have now been
reported in the few HST/STIS echelle spectra with
high S/N \cite{rss+06,lsr+07}.  Because these detections lie
near the sensitivity limit of the spectra, it is not possible
to exclude the possibility that these broad features arise from the blends
of several weak, narrow absorbers.  Furthermore, independent
analysis of these same datasets have questioned
most of the putative detections (Stocke, priv.\ comm.; Danforth \& Shull, in preparation).
These issues can only be resolved by higher S/N observations
and, ideally, analysis of the corresponding \lyb\ profiles.
We look forward to the installation of HST/COS which should
enable such observations.
Even if one can conclusively demonstrate the presence of warm/hot
gas with broad \lya\ absorbers, there are limitations to directly
assessing the mass density of this gas from empirical observation.
First, studies of numerical simulations reveal that
broad \lya\ lines arise from gas whose kinematics are
a mixture of thermally and turbulently broadened motions making
it difficult to derive the temperature and obtain an ionization
correction \cite{rfb06}.
Second, these same models predict that photoionization
processes are significant for some gas with $T< 10^6$K and this
contribution to the ionization correction
cannot be assessed from HI observations alone.
Finally, the detection of broad \lya\ features becomes prohibitive
for gas with $T > 10^{6.5}$K (i.e.\ $\tau_0 < 0.05$ for $N_H = 10^{20} \cm{-2}$),
which may account for the majority of WHIM material by
mass (e.g. \cite{daveetal01}).
In short, one is driven to statistical
comparisons between numerical simulations and observations of broad
\lya\ absorption and, ultimately, may
be limited to a minor fraction of the WHIM.

At $T>10^{6.3}$K, the ion fraction of O$^{+5}$ is negligible and
broad \lya\ lines are generally too weak to detect with previous
or planned UV spectrometers.  To more efficiently trace this gas,
we must turn to higher ionization states of oxygen and other elements.
One option is the NeVIII doublet which probes gas to $T \approx 10^7$K
and is accessible with UV spectrometers for $z>0.25$ \cite{savage05}.
The most sensitive tracers of $T>10^6$K gas, however, are resonance
lines of OVII (21.602 \AA), OVIII (18.969 \AA), and the NeVIII doublet (770,780 \AA).
While NeVIII may be detectable at $z > 0.5$ with COS, for
current X-ray satellites the oxygen lines are a demanding observational
challenge; one is limited by both spectral resolution and instrument
collecting area.  Nicastro and collaborators
have addressed the latter issue (in part) by monitoring X-ray variable
sources that show occasional, bright flares.  During a flare event,
one obtains
target-of-opportunity observations to yield a relatively high S/N dataset
\cite{fangetal2002,nme+05,wmn06}.
While this program has been successful observationally,
the purported detections of
intergalactic OVII and OVIII remain controversial \cite{kwh06,rkp+07}.
While we are optimistic that these programs will reveal a few
indisputable detections before the demise of the Chandra and XMM observatories,
it is evident that a statistical survey for OVII and OVIII awaits
future instrumentation (e.g.\ Con-X, XEUS).
Because OVI may indicate cooling gas within a hot medium
(or hot-cold interfaces), the numerous detections of OVI can serve
as signposts to the `true' WHIM \citep{shulletal03}.

Ultimately, however, the challenge will remain to convert these
tracers of hot gas into constraints on the mass density and spatial
distribution of WHIM gas.  At a minimum, this requires assumptions
about the metal-enrichment of the intergalactic medium.

In summary, we find the current
empirical evidence for the WHIM to be intriguing but far from definitive.
One must consider additional means for exploring
the WHIM that would be enabled by new, proposed or planned instrumentation.
One of these is to explore the line {\it emission} (e.g.\ \lya, OVI, OVII)
from the denser WHIM gas \cite{cf06,bsd08}, which could be accomplished
with UV and X-ray missions currently in design (e.g.\ {\it IGM,XENIA}).
If combined with absorption-line studies, one may map out the mass density
and spatial distribution of this gas at low $z$.
Finally, it may be possible to cross-correlate the Sunyaev-Zeldovich
(SZ) signal of the CMB
with WHIM structures to infer the density and spatial distribution
of this ionized gas \cite{hhp+08}.
This might be accomplished with currently planned SZ experiments.

\begin{figure}
\includegraphics[scale=0.7,angle=0]{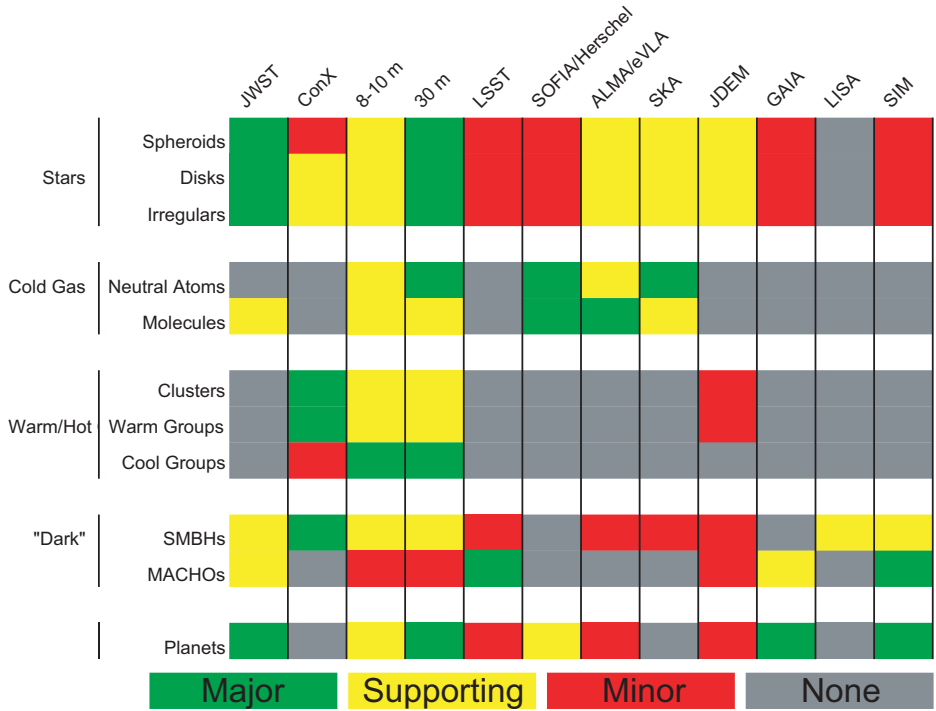}
\caption{For specific missions or categories of facility, the colors represent the role they are expected to play in measuring the budget of baryons in each phase discussed in the text. }\label{fig:missions}
\end{figure}

\section{Future Facilities and the Baryon Census}

In the spirit of the meeting ``Astrophysics in the Next Decade: JWST and Concurrent Facilities'', we consider how the baryon census will be advanced by the facilities in development now for operations in the next decade. To stimulate discussion at the conference, we composed a chart showing the applicability of each mission or facility to the important reservoirs of baryonic matter. The list of missions under consideration was obtained from the conference website. This chart appears in Figure~\ref{fig:missions}, with the contributions of each mission color coded for major (green), supporting (yellow), minor (red), or no (grey) role. These assignments were made in the idiosyncratic judgment of one of us (J. T.) and should not be taken too literally. Each mission has its own, presumably compelling, scientific justification that in most cases does not explicitly include the baryon census (though JWST and Con-X are notable exceptions). The assignments were based primarily on wavelength coverage, spectral and spectroscopic resolution, field of view, and a review of the mission's public science case where available.

The overlap between missions and baryonic components raise some interesting points. First, the future looks relatively bright for the baryon census as a whole. JWST, Con-X, and ALMA are the three most important missions for resolving uncertainties in the census of baryons in stars, hot gas, and cold gas respectively. These are the areas of greatest uncertainty today, so this is somewhat
encouraging for the future. However, only two of these facilities are in development phases: Con-X is still under study and could not reasonably be expected to fly until late in the next decade or after 2020.

The ``Assembly of Galaxies'' is one of the four top-level science drivers of JWST, which will have the sensitivity and wavelength coverage to improve on stellar mass estimates at high redshift and the ability to measure IMFs by star counts over much larger local volumes than has been possible to date. As this is the most uncertain step in the derivation of the cosmic stellar mass budget, assessing the variation of the IMF from place to place and over time should be a primary goal of galaxy formation and evolution studies with JWST.

It looks as though the next decade will bring major advances in the study of the cold neutral and molecular phases of baryons over a wide redshift range, thanks to ALMA and eVLA. As mentioned above, the greatest need in this area is for a more thorough testing of the conversion factor from CO mass to H$_2$ mass in a broad range of astrophysical environments.

One pleasant surprise from this exercise is the notable versatility of large, ground-based optical telescopes (Keck and VLT today, and the $20 - 30$ m ELTs in the future). Though they can lead the way for direct studies of stellar masses over a wide range of redshifts, they also support investigations whose primary use is in another wavelength range. Often this role requires them to obtain supporting data such as spectroscopic redshifts for high-$z$ galaxies, optical counterparts for X-ray or radio sources, and the measurement of dynamical masses by high-resolution studies of linewidths in absorption or emission. It is likely that this versatility reflects the long history of optical astronomy, where the most mature technologies and techniques are available, and their widespread availability compared with the scarce resource of space-based observing time.

This exercise also highlights a broad trend we may expect from astrophysics in the next decade. Since the baryon census encompasses many diverse astrophysical environments, from stars to the diffuse IGM, and requires a wide range of observational techniques, the capability of a facility to address the baryon census reflects its ability to address astrophysical problems in general. With this in mind, we note that many of the space missions in question are tailored to specific scientific questions rather than a broad, community-generated scientific program. While it is difficult to see how a mission like LISA could have collateral uses beyond the detection of gravity waves from a few astrophysical sources, even missions carrying more conventional instruments like JDEM, GAIA, and SIM are fine-tuned to specific scientific cases. This trend toward specialization, even among flagship-scale missions, will evidently be a hallmark of astrophysics in the next decade.

Finally, we note that perhaps the greatest need for new capabilities to address the ``missing baryons'' lies in the X-ray, and that there is no realistic chance for Con-X or something like it to launch before the very end of the next decade.
Thus it is possible that astronomers will, by 2015, see the ``first galaxies'' (JWST), measure DE equation of state (LSST+JDEM), weigh molecules at $z \sim 6$ (ALMA), discover gravitational waves (LISA), find earthlike extrasolar planets (SIM+GAIA), and still not know where all the ordinary matter is or what phase it is in.
This long interval would place the solution to the missing baryons problem at least 20 years away from when it was first recognized. This may be one of those cases where the urgency of performing an observation is tempered by the conventional wisdom that we already know the answer  (the WHIM). But theory has been wrong before, and the only way to know is to do the experiment. If the last decade of astrophysics is a reliable guide, the next ten years will bring a few surprises. The what, when, and where of the baryons may well be among them.

\begin{acknowledgement}
We acknowledge helpful criticism from M. Shull.
J. X. P. is partially supported by NASA/Swift grant
NNX07AE94G and an NSF CAREER grant (AST-0548180).
J. T. gratefully acknowledges the generous support of Gilbert and Jaylee
Mead for their namesake fellowship in the Yale Center for Astronomy and Astrophysics.
\end{acknowledgement}


\end{document}